\documentstyle[epsfig,12pt]{article}
\newcommand{\ndt}{\noindent}

\begin{document}

\large
\begin{flushright}
{\bf DFUB 2000-9} \par
{\bf Bologna, May 2000}
\end{flushright}

\vspace{2cm}

\begin{center}
\Large{\bf Magnetic Monopole Bibliography}
\end{center}

\vspace{2cm}

\begin{center}
{\bf \boldmath 
G. Giacomelli$^{1}$, M. Giorgini$^{1}$, T. Lari$^{1}$, M. Ouchrif$^{~1,2}$,
\par L. Patrizii$^{1}$, V. Popa$^{1,3}$, P. Spada$^{1}$ and V. Togo$^{1}$}
\end{center} 

\begin{center}
{\it $^1$Dipartimento di Fisica dell'Universit\`a di Bologna and INFN, Sezione
di Bologna, I-40127 Bologna, Italy \\
$^2$Faculty of Sciences, University Mohamed I, B.P. 524 Oujda, Morocco \\
$^3$Institute of Space Sciences, Bucharest R-76900, Romania}
\end{center}

\vspace{1.5cm}
\begin{abstract}
We present a bibliography compilation  on magnetic 
monopoles updated to include references till 
the end of year 1999. It
is intended to contain nearly all the experimental papers on the
subject and only the theoretical papers which have some specific experimental
implications.
\end{abstract}

\newpage
\normalsize

{\Large \bf 1. Introduction}
\vskip 15 pt

Even though Maxwell's equations formally allow the existence of 
magnetic monopoles (MMs), interest in this kind of objects arose only
in 1931 after
the paper of P. A. M. Dirac [31D1], in which it was 
shown that magnetic charges can be 
introduced in the framework of quantum mechanics and that the product of the 
basic electric charge and of the basic magnetic charge is quantized
according to the Dirac relation 
$eg = \frac{n\hbar c}{2}$, where $n$ is an integer. Such a particle 
is called  {\em magnetic monopole} 
if it carries only a magnetic charge, and
{\em dyon}
if it carries both magnetic and electric charges (a monopole bound
with a nucleus behaves effectively as a dyon). Dirac could not constrain the
monopole mass; rough estimates indicated that the MM mass should be larger than 
several GeV. 

Many types of searches for magnetic monopoles with 
masses not much larger 
than the proton mass were performed at each new accelerator and in bulk 
matter. Very many 
theoretical studies on MMs have been published. 

The other date of fundamental importance in the history of monopoles is 1974.
In that year 't Hooft [74H1] and Polyakov [74P1] demonstrated that 
Grand Unified Theories (GUT) of the electroweak and strong interactions
implied the existence of MMs with masses of the order of 
$10^{17}\mbox{ GeV/c}^{2}$ ($m_{M} \sim m_{X}/\alpha$ where $m_{X}$ is the 
mass of the carrier of the unified force and $\alpha$ is the unified coupling 
constant). These masses are too large for monopoles to be 
produced at present or future high energy accelerators or somewhere in the 
present universe. They could have been  
produced immediately after  the big bang, either as topological defects or in 
very high energy collisions such as $e^{+}e^{-} \rightarrow M\overline{M}$, 
immediately after the phase transition at the end of the GUT epoch; 
thus GUT monopoles could be present in the cosmic radiation,
since the lightest monopole should be stable, due to 
conservation of magnetic charge. 
From 1974 to the present time a very large number of theoretical 
studies were made on magnetic monopoles; also many experimental searches were 
performed.

The present paper gives a bibliography of publications on monopoles. The 
bibliography is intended to contain nearly all the experimental 
papers on the 
subject and only the theoretical papers which have 
specific experimental implications.
With some exceptions only papers published in international refereed journals 
have
been included.

The publications on MMs comprise many different subjects.

1) Theoretical works on Dirac MMs (for ex. [31D1], [66S1], [76W1], [77K2], 
[97I2], ...). 
The possible existence of bound states of 
magnetic monopoles with electrons and/or nuclei has also been investigated
(for ex. [51M1], [77K3], [83B1], [83B2], [84B1], ...).

2) Theoretical works on GUT MMs (for ex. [74H1], [74P1], [84P1], ...). 

3) Papers on the catalysis of baryon decay 
by GUT monopoles, such as [80R1], [82C2].

4) Papers on the cosmological production of MMs (for ex. 
[76K1], [79P1], [80L2], [80E1], [81G2], ...)  and papers which derived limits 
on MM fluxes from astrophysical considerations (for ex. [70P1],
[82K2], [85A2],
[85B2], ...).

5) Theoretical works on supermassive MMs based on other 
theories, like monopoles from superstrings ([87L1]), 
 intermediate symmetry breaking monopoles ([83L1], [84L2]), 
 or lighter monopoles of electroweak nature ([97C1], [97C4]).  

6) Studies of the energy losses of monopoles in matter and on the possible 
techniques to detect them (for ex. [78A1], [82A4], [83A2], [83D3], [84B2],
[85B5], [87F1], [89P1], [91O1],
[97A1], ...). 

7) Searches at high energy accelerators; the searches are 
either {\em direct} (detection of monopoles 
immediately after their production in high-energy collisions) or 
{\em indirect} (for example when a piece of matter is exposed to
a beam for a long time, and then later analyzed). 
Examples of direct searches are [75G1], [82K1], [83A7], [83M1], [87G2], 
[90B3], [00B2], ...
Examples of indirect searches  are [61F1], [63A1], [66A1], 
[74C1], [75E1], [78C1], [83B7], ....

8) Searches for possible effects of virtual monopoles, such as 
[95D2], [95A2], [97K2], ...

9) Direct searches for MMs in the cosmic radiation. Examples of such searches 
are those by MACRO 
[91B1], [91P1], [94A1],
[95A1], [95M2], [97A2], [97A3] and by 
other experiments [82B2], [83C1], [86P1], [90B1], [90B2], 
[90B4], [90G1], [91O1], [92T1].

10) Indirect searches for monopoles in the cosmic radiation; 
the experiments look for monopoles trapped 
in matter or for the effects due to the passage of MMs in the past. 
Examples are [63G2], [73R1], [83E1], [84P3], [86P1], [87E1], [89A1], [90G1],
[95J1], ...

11) Review papers on various aspects of monopoles and on their 
experimental search; examples are  
[81G3], [82G1], [83P1], [86G2], [94G1], [96B3], ...

Several MM bibliographies have been made in the past, see 
[73S1], [77C1], [80R1], [82C4], [84G1], [94G1]. 
The bibliography in [84G1] is more complete than the present one for the period 
before 1984. The bibliography of [80R1] covers essentially all the papers 
dealing with classical Dirac monopoles.

A bibliography on the experimental limits and on astrophysical bounds  is 
presented regularly every two years by the Particle Data Group (see 
[96P1], [94P2], [92P1]).

The present bibliography covers the period before May 2000 and
is an update of DFUB 98-9 [98G6]. 

Fig.~1a presents an hystogram with the number of papers on MMs
and dyons published each year from 
1973 to 1983 [83P1]; Fig.~1b shows the number of papers in the 
present bibliography 
(until December 1999) as a function of the year of 
publication; Fig.~1c shows an hystogram of the number of yearly papers 
in the SLAC database which meet one of the following conditions:
\newline
- they have {\em monopole} or {\em monopoles} in the title;
\newline
- they have {\em dyon} or {\em dyons} or {\em dyonic} in the title;
\newline
- they have the keyword {\em magnetic monopole} assigned. 

For the first and second hystogram the year is that of publication, 
for the third it is the 
year of receipt by the SLAC library;  
 this third hystogram has many more entries, mostly theoretical 
papers, conference proceedings and unpublished reports. 

The peak rates in the hystograms correspond 
to the periods immediately after the paper of 't Hooft and Polyakov 
(1974) and 
after the 1982 Cabrera candidate event.  

Fig.~2 shows a compilation of 
the  90\% C.L. limits on MMs in the cosmic radiation;
the limits apply to an isotropic flux of bare $ g=g_{D} $ massive magnetic 
monopoles for a catalysis cross section smaller than few mb.

\vskip 15 pt

{\small In the literature one finds references to many types of MMs. For
completeness we recall the simplest definitions of many of these MMs.\par
\ndt $-$ {\it GUT monopoles} are the MMs connected with the 
Grand Unification of the
electroweak and strong interactions and have masses of $10^{16} \div 10^{17}$
GeV. They appear in the early universe at the end of the GUT epoch 
[74H1], [74P1], [83F2], [84P3], [85K2]; {\it 't Hooft-Polyakov monopoles} 
are GUT MMs [90K1], [92B1], [93D1], [98K1]; {\it SO(3)-Z2 monopoles} are 
particular GUT MMs [98G4]. \par\ndt
$-$ {\it QCD monopoles} are MMs with a colour charge [98S1], [99G1]. \par\ndt
$-$ {\it BPS monopoles} are MMs appearing in the Bogomolny and 
Prasad-Sommerfield limit [93B2], [97B4], [97C3], [97S1], 
 [98B2], [99I2], [99L1]. \par\ndt
$-$ {\it Kaluza-Klein monopoles} are MMs connected with the unification 
of the GUT interaction with the gravitational interaction; they have typical
masses $> 10^{19}$ GeV [83S3], [86S3], [97B3], [98B3]. \par\ndt
$-$ {\it Non abelian monopoles} are MMs which appear in non abelian theories
(including GUT, Kaluza-Klein,..., monopoles) 
[94M1], [94M2], [99C1], [00L1]. \par\ndt
$-$ {\it Classical (Dirac) monopoles} are abelian MMs hypothesized by Dirac
in 1931; they could have relatively low masses [31D1], [33T1], [51M1], 
 [59B1], [72B2], [78Z1], [90A4], [97H1]. \par\ndt
$-$ {\it Intermediate mass monopoles} are MMs connected with an intermediate
symmetry breaking scale; they could have masses of the order of $10^{10}$ 
GeV [84L2], [95D2], [00B1], [00B2]. \par\ndt
$-$ {\it Wu-Yang monopoles} are particular solutions in Yang-Mills 
theories [76W1], [98D2]. \par\ndt
$-$ {\it Monopoles in String Theories} [93S1], [94B2], [99H1]. \par\ndt
$-$ {\it Constituent monopoles} are MMs formed by the superposition of
$n$ BPS MMs (for a SU(n) gauge group) [98K2], [99K1]. \par\ndt
$-$ {\it Complex monopoles} are solutions of Yang-Mills theories when 
adding a Chern-Simons term [98T1], [99H2], [99T1]. \par\ndt
For further possibilities, see other references, such as 
[98L2], [99F1], [99T2]. \par 
}

\vskip 15 pt

We gratefully acknowledge many members of the MACRO Collaboration, in 
particular all the members of the Bologna group. We apologize for 
possible omissions.

\newpage
\vskip 15 pt
{\Large \bf 2. References on Magnetic Monopoles}
\vskip 15 pt

\begin{description}

\small

\item{[31D1]} P. A. M. Dirac (Quantized singularities in the electromagnetic
field) Proc. Roy. Soc. Lond. 133 (1931) 60. 
\vspace{-3.5 mm}
\item{[31T1]} J. Tamm (Die verallgemeinerten kugelfunktionen und die 
wellenfunktionen eines elektrons un felde eines magnetpoles) 
Z. Phys. 71 (1931) 141. 

\item{[33T1]} M. A. Tuve (Search by deflection-experiments for the Dirac 
isolated magnetic pole) Phys. Rev. 43 (1933) 770. 

\item{[35G1]} B. O. Gr\"{o}nblom (\"{U}ber singul\"{a}re magnetpole) 
Z. Phys. 98 (1935) 283. 

\item{[44E1]} F. Ehrenhaft (New experiments about the magnetic current) 
Phys. Rev. 65 (1944) 62.

\item{[45E1]} F. Ehrenhaft (The measurements of single magnetic charges 
and the electrostatic field around the permanent magnet) 
Phys. Rev. 67 (1948) 201.

\item{[48D1]} P. A. M. Dirac (The theory of magnetic poles) Phys. Rev. 74
(1948) 817.

\item{[51M1]} W. Malkus (The interaction of the Dirac magnetic monopole
with matter) Phys. Rev. 83 (1951) 899. 

\item{[58G1]} E. Goto ( On the observation of magnetic poles
) J. Phys. Soc. Jpn. 13 (1958) 1413.

\item{[59B1]} H. Bradner and W. M. Ishell (Search for Dirac monopoles) 
Phys. Rev. 114 (1959) 603.

\item{[61F1]} M. Fidecaro et al. (Search for magnetic monopoles)
Nuovo Cim. 22 (1961) 657.

\item{[62C1]} N. Cabibbo and E. Ferrari (Quantum electrodynamics with Dirac
monopoles) Nuovo Cim. 23 (1962) 1147. 

\item{[63A1]} E. Amaldi et al. (Search for Dirac magnetic poles)
Nuovo Cimento 28 (1963) 773.
\vspace{-3.5 mm}
\item{[63G1]} E. Goto (Expected behaviour of the Dirac monopole in the cosmic 
space) Prog. Theor. Phys. 30 (1963) 700.
\vspace{-3.5 mm}
\item{[63G2]} E. Goto et al. (Search for ferromagnetically trapped magnetic 
monopoles of cosmic-ray origin) Phys. Rev. 132 (1963) 387.
\vspace{-3.5 mm}
\item{[63P1]} E. M. Purcell et al. (Search for the Dirac monopole with 30-BeV
protons) Phys. Rev. 129 (1963) 2326.

\item{[65C1]} R. A. Carrigan (Consequences of the existence of massive 
magnetic poles) Nuovo Cimento 38 (1965) 638.
\vspace{-3.5 mm}
\item{[65T1]} D. R. Tompkins (Total energy loss and \v{C}erenkov emission from 
monopoles) Phys. Rev. 138 (1965) 248. 
\vspace{-3.5 mm}
\item{[65W1]} S. Weinberg (Photons and gravitons in perturbative theory: 
derivation of Maxwell's and Einstein's equations) Phys. Rev. 138 (1965) 988.

\item{[66A1]} R. K. Adair et al. (Search for heavy magnetic monopoles) 
Phys. Rev. 149 (1966) 1070.
\vspace{-3.5 mm}
\item{[66S1]} J. Schwinger (Magnetic charge and quantum field theory)
Phys. Rev. 144 (1966) 1087.

\item{[68S1]} J. Schwinger (Sources and magnetic charge)
Phys. Rev. 173 (1968) 1536.

\item{[70C1]} A. Crispin and G. N. Fowler (Density effect in the ionization 
energy loss of fast charged particles in matter) Rev. Mod. Phys. 42 (1970) 290. 
\vspace{-3.5 mm}
\item{[70P1]} E. N. Parker (The origin of magnetic fields) Astrophys. J. 160
(1970) 383. 
\vspace{-3.5 mm}
\item{[70S1]} D. Sivers (Possible binding of a magnetic monopole to a particle
with electric charge and a magnetic dipole moment) Phys. Rev. D2 (1970) 2048.

\item{[71C1]} R. A. Carrigan et al. (Upper limit for magnetic monopole
production by neutrinos) Phys. Rev. D3 (1971) 56. 
\vspace{-3.5 mm}
\item{[71E1]} P. H. Ebherard (Search for magnetic monopoles in lunar material)
Phys. Rev. D4 (1971) 3260.
\vspace{-3.5 mm}
\item{[71K1]} H. H. Kolm et al. (Search for magnetic monopoles)
Phys. Rev. D4 (1971) 1285.
\vspace{-3.5 mm}
\item{[71P1]} E. N. Parker (The generation of magnetic fields in astrophysical 
bodies. The galactic field) Astrophys. J. 163 (1971) 255.

\item{[72B1]} D. F. Bartlett and M. D. Lahana (Search for tachion monopoles)
Phys. Rev. D6 (1972) 1817.
\vspace{-3.5 mm}
\item{[72B2]} L. M. Barkov et al. (Search for Dirac monopoles in the 70-BeV 
proton syncrotron) Sov. Phys. JETP 34 (1972) 917.
\vspace{-3.5 mm}
\item{[72M1]} V. P. Martem'yanov and S. K. Khakimov (Slowing down of a Dirac 
monopole in metals and ferromagnetic substances) Sov. Phys. JETP 35 (1972) 20.

\item{[73C1]} R. A. Carrigan et al. (Search for magnetic monopole production by
300-GeV protons) Phys. Rev. D8 (1973) 3717. 
\vspace{-3.5 mm}
\item{[73R1]} R. R. Ross et al. (Search for magnetic monopoles in lunar 
material using an electromagnetic detector) Phys. Rev. D8 (1973) 698.
\vspace{-3.5 mm}
\item{[73S1]} D. M. Stevens (Magnetic monopoles: an updated bibliography)
UPI-EPP-735 (1973).

\item{[74C1]} R. A. Carrigan et al. (Extension of Fermi National Accelerator
Laboratory magnetic monopole search to 400-GeV) Phys. Rev. D10 (1974) 3867. 
\vspace{-3.5 mm}
\item{[74H1]} G. 't Hooft (Magnetic monopoles in unified gauge theories)
Nucl. Phys. B29 (1974) 276. 
\vspace{-3.5 mm}
\item{[74H2]} R. Howard (Studies of solar magnetic fields)
Solar Phys. 38 (1974) 283.
\vspace{-3.5 mm}
\item{[74P1]} A. M. Polyakov (Particle spectrum in quantum field theory)
JETP Lett. 20 (1974) 194.

\item{[75B1]} A. P. Balachandran et al. (Monopole theories with massless
and massive gauge fields) Phys. Rev. D11 (1975) 2260.  
\vspace{-3.5 mm}
\item{[75C1]} R. A. Carrigan Jr. and F. A. Nezrick (Search for neutrino-produced
magnetic monopoles in a bubble chamber exposure) Nucl. Phys. B91 (1975) 279.
\vspace{-3.5 mm}
\item{[75E1]} P. H. Ebherard et al. (Evidence at the 1/$10^{18}$ probability 
level against the production of magnetic monopoles in protons interactions at 
300 GeV/c) Phys. ReV. D11 (1975) 3099.
\vspace{-3.5 mm}
\item{[75F1]} M. W. Friedlander (Comments on the reported observation of a
monopole) Phys. Rev. Lett. 35 (1975) 1167. 
\vspace{-3.5 mm}
\item{[75F2]} R. L. Fleischer et al. (Probabilities for an alternative
explanation of the moving magnetic monopole) Phys. Rev. Lett. 35 (1975) 1412. 
\vspace{-3.5 mm}
\item{[75G1]} G. Giacomelli et al. (Search for magnetic monopoles at the
 CERN-ISR with plastic detectors) Nuovo Cim. 28A (1975) 21.
\vspace{-3.5 mm}
\item{[75H1]} E. V. Hungerford (Comment on the observation of a moving magnetic 
monopole) Phys. Rev. Lett. 35 (1975) 1303.
\vspace{-3.5 mm}
\item{[75J1]} B. Julia and A. Zee (Poles with both magnetic and electric charges
in non-abelian gauge theories) Phys. Rev. D11 (1975) 2227.
\vspace{-3.5 mm}
\item{[75M1]} R. Mignani and E. Recami (Connection between magnetic monopoles  
and faster-than-light speeds: answer to the comments by Corben and Honig) 
Lett. Nuovo Cimento 13 (1975) 589.
\vspace{-3.5 mm}
\item{[75P1]} P. B. Price et al. (Evidence for detection of a moving magnetic
monopole) Phys. Rev. Lett. 35 (1975) 487. 

\item{[76A1]} S. P. Ahlen (Monopole track characteristics in plastic 
detectors) Phys. Rev. D14 (1976) 2935. 
\vspace{-3.5 mm}
\item{[76C1]} R. A. Carrigan et al. (Search for misplaced magnetic monopoles) 
Phys. Rev. D13 (1976) 1823.
\vspace{-3.5 mm}
\item{[76C2]} J. M. Cornwall et al. (Relation between monopole mass and 
primary monopole flux) Phys. Rev. Lett. 36 (1976) 900. 
\vspace{-3.5 mm}
\item{[76G1]} A. S. Goldhaber (Spin and statistics connection for charge -
monopole composites) Phys. Rev. Lett. 36 (1976) 1122.
\vspace{-3.5 mm}
\item{[76K1]} T. W. B. Kibble (Topology of cosmic domains and strings) 
J. Phys. A9 (1976) 1387.
\vspace{-3.5 mm}
\item{[76W1]} T. T. Wu and C. N. Yang (Dirac monopole without strings:
monopole harmonics) Nucl. Phys. B107 (1976) 365.

\item{[77B1]} R. A. Brandt et al. (Dirac monopole theory with and without
strings) Phys. Rev. D15 (1977) 1175. 
\vspace{-3.5 mm}
\item{[77C1]} R. A. Carrigan (Magnetic monopole bibliography: 1973-1976)
Fermilab 77/42 (1977) 130.
\vspace{-3.5 mm}
\item{[77F1]} P. M. Fishbane et al. (Diffractive inelastic monopole
transitions and the slope - mass relation in $\pi$N production) Phys.
Rev. D15 (1977) 782. 
\vspace{-3.5 mm}
\item{[77K1]} A. Kupiainen et al. (On the uniqueness of monopole solution)
Phys. Lett. B67 (1977) 80.
\vspace{-3.5 mm}
\item{[77K2]} Y. Kazama et al. (Scattering of a Dirac particle with charge 
Z$\ast e$ by a fixed magnetic monopole) Phys. Rev. D15 (1977) 2287. 
\vspace{-3.5 mm}
\item{[77K3]} Y. Kazama et al. (Existence of bound states for a charged
 spin-$\frac{1}{2}$ particle with an extra magnetic moment in the field
of a fixed magnetic monopole) Phys. Rev. D15 (1977) 2300. 
\vspace{-3.5 mm}
\item{[77W1]} T. T. Wu (Some properties of monopole harmonics) Phys. Rev. D16
(1977) 1018.

\item{[78A1]} S. P. Ahlen (Stopping power formula for magnetic monopoles)
Phys. Rev. D17 (1978) 229. 
\vspace{-3.5 mm}
\item{[78C1]} R. A. Carrigan et al. (Search for magnetic monopoles at the
CERN Intersecting Storage Rings) Phys. Rev. D17 (1978) 1754.
\vspace{-3.5 mm}
\item{[78C2]} P. H. Cox et al. (Bound states with a gauge monopole) Phys.
Rev. D18 (1978) 1211. 
\vspace{-3.5 mm}
\item{[78D1]} P. A. M. Dirac (The monopole concept) Int. Jour. of Theor. Phys.
17 (1978) 235.  
\vspace{-3.5 mm}
\item{[78G1]} G. Giacomelli (Searches for missing particles) invited paper at
The 1978 Singapore Meeting on Frontiers of Physics, Proceedings of the 
Conference (1978). 
\vspace{-3.5 mm}
\item{[78P1]} P. B. Price et al. (Further measurements and reassessment of the 
magnetic-monopole candidate)
Phys. Rev. D18 (1978) 1382. 
\vspace{-3.5 mm}
\item{[78Z1]} V. P. Zrelov (On possible improvement of photographic detector
to search for Dirac monopole by Vavilov-\v{C}erenkov radiation) Nucl. Instr.
Meth. 153 (1978) 145.  
\vspace{-3.5 mm}
\item{[78Z2]} Y. B. Zeldovich et al. (On the concentration of relic magnetic 
monopoles in the Universe) Phys. Lett. 79B (1978) 239.

\item{[79B1]} J. J. Broderick et al. (Observational limits on the magnetic
monopole structure of protons) Phys. Rev. D19 (1979) 1046. 
\vspace{-3.5 mm}
\item{[79F1]} I. M. Frank (Transition radiation of the magnetic charge)
Sov. J. Nucl. Phys. 29 (1979) 90. \par 
\vspace{-3.5 mm}
\item{[79K1]} R. Kerner (Energy levels of the magnetic monopole in the 
Prasad-Sommerfield limit) Phys. Rev. D19 (1979) 1243.
\vspace{-3.5 mm}
\item{[79N1]} W. Nahm (Interacting monopoles) Phys. Lett. B85 (1979)
373.
\vspace{-3.5 mm}
\item{[79O1]} L. O'Raifeartaigh et al. (On magnetic monopole interactions)
Phys. Rev. D20 (1979) 1941.
\vspace{-3.5 mm}
\item{[79P1]} J. P. Preskill (Cosmological production of superheavy magnetic
monopoles) Phys. Rev. Lett. 43 (1979) 1365.
\vspace{-3.5 mm}
\item{[79R1]} H. M. Ruck et al. (Comment on the motion of a spin $\frac{1}{2}$
particle in the field of a magnetic monopole) Phys. Rev. D20 (1979) 2089. 
\vspace{-3.5 mm}
\item{[79W1]} E. Witten (Dyons of charge $e\, \theta$ / 2 $\pi$) Phys. Lett. B86
(1979) 283.

\item{[80A1]} S. P. Ahlen (Theoretical and experimental aspects of the energy
loss of relativistic heavily ionizing particles) Rev. Mod. Phys. 52 (1980) 121. 
\vspace{-3.5 mm}
\item{[80B1]} F. A. Bais et al. (On the suppression of monopole production in
the very early universe) Nucl. Phys. B170 (1980) 507.
\vspace{-3.5 mm}
\item{[80C1]} Y. M. Cho (Colored monopoles) Phys. Rev. Lett. 44 (1980) 1115,
erratum-ibid 44 (1980) 1566.
\vspace{-3.5 mm}
\item{[80C2]} R. A. Carrigan jr. (Grand unification magnetic monopoles inside 
the Earth) Nature (London) 288 (1980) 348.
\vspace{-3.5 mm}
\item{[80D1]} C. P. Dokos et al. (Monopoles and dyons in the SU(5) model)
Phys. Rev. D21 (1980) 2940. \par
\vspace{-3.5 mm}
\item{[80E1]} M. B. Einhorn et al. (Are Grand Unification Theories compatible 
with standard cosmology?) Phys. Rev. D21 (1980) 3295.
\vspace{-3.5 mm}
\item{[80F1]} D. Fryberger (On the magnetically bound monopole pair, a 
possible structure for fermions) Nuovo Cim. Lett. 28 (1980) 313.
\vspace{-3.5 mm}
\item{[80K1]} T. W. B. Kibble (Some implication of a cosmological phase 
transition) Phys. Rept. 67 (1980) 183.
\vspace{-3.5 mm}
\item{[80L1]} G. Lazarides and Q. Shafi (The fate of primordial magnetic
monopoles) Phys. Lett. B94 (1980) 149.
\vspace{-3.5 mm}
\item{[80L2]} P. Langacker and S. Y. Pi (Magnetic
monopoles in Grand Unified Theories) Phys. Rev. Lett. 45 (1980) 1.
\vspace{-3.5 mm}
\item{[80R1]} J. Ruzicka and V. P. Zrelov (Fifty years of Dirac monopole:
complete bibliography) JINR-1-2-80-850 (1980).
\vspace{-3.5 mm}
\item{[80S1]} D. M. Scott (The masses of monopoles in Grand Unified Theories)
Nucl. Phys. B171 (1980) 109. 

\item{[81B1]} J. Burzlaff (On the SU(3) monopole with magnetic quantum numbers
(0,2)) Phys. Rev. D23 (1981) 1329. 
\vspace{-3.5 mm}
\item{[81B2]} R. A. Brandt et al. (Magnetic monopoles in SU(N) gauge theories)
Nucl. Phys. B186 (1981) 84. 
\vspace{-3.5 mm}
\item{[81B3]} J. D. Bowman et al. (Pion charge exchange reaction as a probe
of isovector monopole resonances) Phys. Rev. Lett. 46 (1981) 1614.
\vspace{-3.5 mm}
\item{[81B4]} D. F. Bartlett et al. (Search for cosmic-ray-related magnetic
monopoles at ground level) Phys. Rev. D24 (1981) 612. 
\vspace{-3.5 mm}
\item{[81C1]} G. Calucci (Capture of nucleons by a monopole) Lett. Nuovo Cim.
32 (1981) 201.
\vspace{-3.5 mm}
\item{[81C2]} G. Calucci (Eikonal formulation for the scattering by a monopole
and by a dyon) Nuovo Cim. Lett. 32 (1981) 205.
\vspace{-3.5 mm}
\item{[81C3]} G. P. Cook et al. (Supercooling in the SU(5) phase transitions
and magnetic monopole suppression) Phys. Rev. D23 (1981) 1321.
\vspace{-3.5 mm}
\item{[81C4]} H. Chan et al. (Monopole charges in unified gauge theories)
Nucl. Phys. B189 (1981) 364. 
\vspace{-3.5 mm}
\item{[81D1]} A. K. Drukier (The creation of magnetic monopoles in outer gaps
of pulsars) Astrophys. Space Sci. 74 (1981) 245.
\vspace{-3.5 mm}
\item{[81E1]} M. B. Einhorn et al. (Monopole production in the very early
universe in a first order phase transition) Nucl. Phys. B180 (1981) 385.
\vspace{-3.5 mm}
\item{[81G1]} P. Goddard et al. (The magnetic charges of stable selfdual
monopoles) Nucl. Phys. B191 (1981) 528.
\vspace{-3.5 mm}
\item{[81G2]} A. H. Guth (Inflationary universe: A possible solution to 
the horizon and flatness problems) Phys. Rev. D23 (1981) 347.
\vspace{-3.5 mm}
\item{[81G3]} G. Giacomelli (Review of the experimental status (past and 
future) of monopole searches) in Proceedings of the Conference on Monopoles 
in Quantum Field Theory, Trieste (1981). Revised for the 1982 Zuoz Spring
School of Physics.
\vspace{-3.5 mm}
\item{[81G4]} G. Giacomelli and G. Kantarjian (Magnetic monopole searches at 
Isabelle) in Proceedings of the 1981 Isabelle Summer Workshop (1981).
\vspace{-3.5 mm}
\item{[81K1]} J. E. Kim (Flavor unity in SU(7): low mass magnetic monopole,
doubly charged lepton and Q=$\frac{5}{3},- \frac{4}{3}$ quarks) Phys. Rev.
D23 (1981) 2706.
\vspace{-3.5 mm}
\item{[81K2]} T. W. Kirkman et al. (Asymptotic analysis of the monopole
structure) Phys. Rev. D24 (1981) 999.
\vspace{-3.5 mm}
\item{[81K3]} K. Kinoshita and P. B. Price (Study of highly ionizing
particles at mountain altitude) Phys. Rev. D24 (1981) 1707.
\vspace{-3.5 mm} 
\item{[81K4]} R. Kerner (Magnetic monopole in the massless limit) Phys. Rev.
D24 (1981) 2336.
\vspace{-3.5 mm}
\item{[81L1]} G. Lazarides et al. (Superheavy magnetic monopole hunt) Phys.
Lett. B100 (1981) 21. 
\vspace{-3.5 mm}
\item{[81R1]} G. A. Ringwood et al. (Monopoles admit spin) Phys. Rev. Lett.
47 (1981) 625. 
\vspace{-3.5 mm}
\item{[81R2]} V. A. Rubakov (Superheavy magnetic monopoles 
and proton decay) JETP Lett.
33 (1981) 644.
\vspace{-3.5 mm}
\item{[81S1]} D. R. Stump (Monopole ionization and the transition from weak
to strong coupling in gauge theories) Phys. Rev. D23 (1981) 972.
\vspace{-3.5 mm}
\item{[81S2]} P. J. Steinhardt (Monopole dissociation in the early universe)
Phys. Rev. D24 (1981) 842.
\vspace{-3.5 mm}
\item{[81U1]} J. D. Ullman (Limits on the flux of slowly moving very massive
particles carrying electric or magnetic charge) Phys. Rev. Lett. 47 (1981)
289.

\item{[82A1]} E. N. Alekseyev et al. (Search for superheavy magnetic monopoles
at the Baksan underground telescope) Lett. Nuovo Cim. 35 (1982) 413.
\vspace{-3.5 mm}
\item{[82A2]} I. K. Affleck et al. (Monopole pair production in a magnetic
field) Nucl. Phys. B194 (1982) 38. 
\vspace{-3.5 mm}
\item{[82A3]} C. W. Akerlof (Limits on the thermoacoustic detectability of 
electric and magnetic charges) Phys. Rev. D26 (1982) 1116.
\vspace{-3.5 mm}
\item{[82A4]} S. P. Ahlen et al. (Calculation of the stopping power of very
low velocity magnetic monopoles) Phys. Rev. D26 (1982) 2347.
\vspace{-3.5 mm}
\item{[82B1]} M. Blagojevic et al. (The infrared problem and radiation effects 
in monopole processes) Nucl. Phys. B198 (1982) 427. 
\vspace{-3.5 mm}
\item{[82B2]} R. Bonarelli et al. (An experimental search for cosmic monopoles)
Phys. Lett. B112 (1982) 100.
\vspace{-3.5 mm}
\item{[82C1]} B. Cabrera (First results from a superconductive detector for
moving magnetic monopoles) Phys. Rev. Lett. 48 (1982) 1378. 
\vspace{-3.5 mm}
\item{[82C2]} C. G. Callan (Disappearing dyons) Phys. Rev. D25 (1982) 2141.
\vspace{-3.5 mm}
\item{[82C3]} C. G. Callan (Dyon-fermion dynamics) Phys. Rev. D26 (1982) 2058.
\vspace{-3.5 mm}
\item{[82C4]} R. E. Craven, W. P. Trower (Magnetic monopole bibliography 
1981-1982) Fermilab-82/96.
\vspace{-3.5 mm}
\item{[82D1]} A. K. Drukier et al. (Monopole pair creation in energetic
collisions: is it possible?) Phys. Rev. Lett. 49 (1982) 102.
\vspace{-3.5 mm}
\item{[82D2]} S. Dimopoulos et al. (Is there a local source of magnetic 
monopoles?) Nature (London) 298 (1982) 824.
\vspace{-3.5 mm}
\item{[82E1]} J. Ellis et al. (Baryon number violation catalyzed by Grand
Unified monopoles) Phys. Lett. B116 (1982) 127.
\vspace{-3.5 mm}
\item{[82F1]} G. W. Ford (Energy loss by slow magnetic monopoles traversing
a metal) Phys. Rev. D26 (1982) 2519.
\vspace{-3.5 mm}
\item{[82G1]} G. Giacomelli (Experimental status of monopoles)
in Proceedings of the Workshop on Magnetic Monopoles (Wingspread, 1982).
\vspace{-3.5 mm}
\item{[82H1]} M. Honda (Magnetic monopole in lattice gauge - Higgs system)
Phys. Lett. B109 (1982) 467. 
\vspace{-3.5 mm}
\item{[82H2]} K. Hayashi (Stopping power for slow monopoles) Nuovo Cim. Lett.
33 (1982) 324. 
\vspace{-3.5 mm}
\item{[82K1]} K. Kinoshita et al. (Search for highly ionizing particles in
$e^+e^-$ collisions at $\sqrt s=29$ GeV) Phys. Rev. Lett. 48 (1982) 77.
\vspace{-3.5 mm}
\item{[82K2]} E. W. Kolb et al. (Monopole catalysis of nucleon decay in neutron
stars) Phys. Rev. Lett. 49 (1982) 1373.
\vspace{-3.5 mm}
\item{[82L1]} G. Lazarides et al. (Consequences of a monopole with Dirac
 magnetic charge) Phys. Rev. Lett. 49 (1982) 1756.
\vspace{-3.5 mm}
\item{[82L2]} M. J. Longo (Massive magnetic monopoles: indirect and direct
limits on their number density and flux) Phys. Rev. D25 (1982) 2399.
\vspace{-3.5 mm}
\item{[82M1]} L. Mizrachi (Comments on 
``dyons of charge $e\, \theta$ / 2 $\pi$")
Phys. Lett. B110 (1982) 242.
\vspace{-3.5 mm}
\item{[82R1]} G. A. Ringwood et al. (Monopoles admit Fermi statistics)
Nucl. Phys. B204 (1982) 168.
\vspace{-3.5 mm}
\item{[82R2]} C. Rubbia (Hunting the supermassive monopole without 
superconductivity) CERN-EP Internal Report 82-01 (1982).
\vspace{-3.5 mm}
\item{[82S1]} A. Soper (Multi - monopoles close together) Nucl. Phys. B199
(1982) 290. 
\vspace{-3.5 mm}
\item{[82T1]} J. S. Trefil (A semiclassical theory for ionization by Grand
Unification magnetic monopoles) Nucl. Phys. B203 (1982) 501.
\vspace{-3.5 mm}
\item{[82T2]} M. S. Turner et al. (Magnetic monopoles and the survival of
 galactic magnetic fields) Phys. Rev. D26 (1982) 1296.
\vspace{-3.5 mm}
\item{[82W1]} E. J. Weinberg (A continuous family of magnetic monopole
solutions) Phys. Lett. B119 (1982) 151. 
\vspace{-3.5 mm}
\item{[82W2]} E. J. Weinberg (Fundamental monopoles in theories with
arbitrary symmetric breaking) Nucl. Phys. B203 (1982) 445. 

\item{[83A1]} N. Auerbach et al. ((P,N) and (N,P) reactions as probes of
 isovector giant monopole resonances) Phys. Rev. C28 (1983) 280.
\vspace{-3.5 mm}
\item{[83A2]} S. P. Ahlen et al. (Can Grand Unification monopoles be detected
with plastic scintillators?) Phys. Rev. D27 (1983) 688.
\vspace{-3.5 mm}
\item{[83A3]} J. Arafune et al. (Limit on the solar monopole abundance)
Phys. Lett. B133 (1983) 380. 
\vspace{-3.5 mm}
\item{[83A4]} S. N. Anderson et al. (Possible evidence for magnetic monopole
 interactions: anomalous long range $\alpha$ particle tracks deep underground)
Phys. Rev. D28 (1983) 2308.
\vspace{-3.5 mm}
\item{[83A5]} J. Arafune and M. Fukugita (Velocity-dependent factors for the
Rubakov process for slowly moving magnetic monopoles in matter) Phys. Rev. Lett.
50 (1983) 1901.
\vspace{-3.5 mm}
\item{[83A6]} S. P. Ahlen et al. (Comment on "searches for slowly moving
magnetic monopoles") Phys. Rev. Lett. 51 (1983) 940. 
\vspace{-3.5 mm}
\item{[83A7]} B. Aubert et al. (Searches for magnetic monopoles in 
proton-antiproton interactions at 540 GeV C.M. energy) Phys. Lett. B120 
(1983) 465.
\vspace{-3.5 mm}
\item{[83A8]} C. W. Akerlof (Intrinsic limits for acoustic detection of 
magnetic monopoles) Phys. Rev. D27 (1983) 1675.
\vspace{-3.5 mm}
\item{[83B1]} L. Bracci and G. Fiorentini (Monopolic atoms and monopole 
catalysis of proton decay) Phys. Lett. B124 (1983) 29.
\vspace{-3.5 mm}
\item{[83B2]} L. Bracci and G. Fiorentini (Binding of magnetic monopoles and
atomic nuclei) Phys. Lett. B124 (1983) 493. 
\vspace{-3.5 mm}
\item{[83B3]} R. Bonarelli et al. (Search for cosmic magnetic monopoles)
Phys. Lett. B126 (1983) 137.
\vspace{-3.5 mm}
\item{[83B4]} M. C. Bowman (A description of E. Weinberg's continuous family
of monopoles using the Atiyah-Drinfeld-Hitchin-Manin-Nahm formalism) Phys.
Lett. B133 (1983) 344. 
\vspace{-3.5 mm}
\item{[83B5]} G. Battistoni et al. (Nucleon stability, magnetic monopoles and 
atmospheric neutrinos in the Mont Blanc experiment) Phys. Lett. B133 (1983)
454. 
\vspace{-3.5 mm}
\item{[83B6]} J. Bartelt et al. (New limit on magnetic monopole flux)
Phys. Rev. Lett. 50 (1983) 655. 
\vspace{-3.5 mm}
\item{[83B7]} S. W. Barwick et al. (Search for penetrating, highly charged
particles at mountain altitude) Phys. Rev. D28 (1983) 2338.
\vspace{-3.5 mm}
\item{[83B8]} A. O. Barut et al. (Capture of a Dirac monopole by a magnetic
dipole) Phys. Rev. D28 (1983) 2666.
\vspace{-3.5 mm}
\item{[83B9]} P. C. Bosetti et al. (Search for magnetic monopoles catalyzing 
barion decay) Phys. Lett. B133 (1983) 265.
\vspace{-3.5 mm}
\item{[83C1]} B. Cabrera et al. (Upper limit on flux of cosmic ray monopoles
obtained with a three-loop superconductive detector) Phys. Rev. Lett. 51 (1983)
1933.
\vspace{-3.5 mm}
\item{[83C2]} B. Cabrera and W. P. Trower (Magnetic monopoles: evidence since 
Dirac's conjecture) Found. Phys. 13 (1983) 1985.
\vspace{-3.5 mm}
\item{[83C3]} G. Callan (Monopole catalysis of baryon decay) Nucl. Phys.
B212 (1983) 391. 
\vspace{-3.5 mm}
\item{[83C4]} M. Claudson et al. (Magnetic monopoles and dyons in N=1
supersymmetric theories) Nucl. Phys. B221 (1983) 461.
\vspace{-3.5 mm}
\item{[83D1]} T. Doke et al. (Search for massive magnetic monopoles using
plastic track detectors) Phys. Lett. 129B (1983) 370.
\vspace{-3.5 mm}
\item{[83D2]} F. Dong et al. (Fractional charges, monopoles and peculiar
photons in SO(18) GUT models) Phys. Lett. B129 (1983) 405.  
\vspace{-3.5 mm}
\item{[83D3]} S. D. Drell et al. (Energy loss of slowly moving magnetic
 monopoles in matter) Phys. Rev. Lett. 50 (1983) 644.
\vspace{-3.5 mm}
\item{[83E1]} T. Ebisu and T. Watanabe (Search for superheavy monopoles in 65 
Kg of iron magnetic sand with a SQUID fluxmeter) J. Phys. Soc. Jpn. 52 (1983) 2617.
\vspace{-3.5 mm}
\item{[83E2]} S. Errede et al. (Experimental limits on magnetic monopole
catalysis of nucleon decay) Phys. Rev. Lett. 51 (1983) 245. 
\vspace{-3.5 mm}
\item{[83E3]} P. Eckert et al. (The mass of the fundamental SU(5) monopole)
Nucl. Phys. B226 (1983) 387.
\vspace{-3.5 mm}
\item{[83F1]} K. Freese et al. (Is the local monopole flux enhanced?) Phys.
Lett. B123 (1983) 293. 
\vspace{-3.5 mm}
\item{[83F2]} D. Fargion (Strong limits on GUT monopole fluxes in magnetic
 universes) Phys. Lett. B127 (1983) 35.
\vspace{-3.5 mm}
\item{[83F3]} K. Freese et al. (Monopole catalysis of nucleon decay in old
pulsars) Phys. Rev. Lett. 51 (1983) 1625. 
\vspace{-3.5 mm}
\item{[83G1]} D. E. Groom et al. (Search for slowly moving massive magnetic
monopoles) Phys. Rev. Lett. 50 (1983) 573. 
\vspace{-3.5 mm}
\item{[83G2]} D. E. Groom et al. (Response to: comment on ``searches for 
slowly moving magnetic monopoles") Phys. Rev. Lett. 51 (1983) 941.
\vspace{-3.5 mm}
\item{[83G3]} A. F. Grillo et al. (Fermion induced monopole - anti-monopole
annihilation) Nuovo Cim. 36 (1983) 579.
\vspace{-3.5 mm}
\item{[83G4]} B. Grossman et al. (The electroweak barrier of the magnetic
monopole) Phys. Rev. D28 (1983) 2109. 
\vspace{-3.5 mm}
\item{[83G5]} G. Giacomelli (Conference high-lights and summation. Experimental
) Invited paper at the Monopole 83 Workshop (Ann Arbor, Mich., 1983).
\vspace{-3.5 mm}
\item{[83K1]} M. Kleber (Molecular ring currents induced by magnetic monopoles)
Z. Phys. 314 (1983) 251.
\vspace{-3.5 mm}
\item{[83L1]} G. Lazarides et al. (Axions and the primordial monopole problem)
Phys. Lett. B124 (1983) 26. 
\vspace{-3.5 mm}
\item{[83L2]} H. J. Lipkin (Effects of magnetic monopoles on nuclear wave 
functions and possible catalysis of nuclear beta decay and spontaneous fission)
Phys. Lett. B133 (1983) 347. \par
\vspace{-3.5 mm}
\item{[83M1]} P. Musset et al. (Search for magnetic monopoles in
 electron-positron collisions at 34 Gev C.M. energy) Phys. Lett. B128 (1983)
333.
\vspace{-3.5 mm}
\item{[83M2]} T. Mashimo et al. (An underground search for anomalous
 penetrating particles such as massive magnetic monopoles) Phys. Lett. B128
(1983) 327.
\vspace{-3.5 mm}
\item{[83M3]} V. F. Mikhailov (The magnetic charge phenomenon in ferromagnetic 
aerosols) Phys. Lett. B130 (1983) 331.
\vspace{-3.5 mm}
\item{[83P1]} R. D. Peccei (Theoretical review of monopole bounds) Munich
preprint MPI-PAE/PTh 83-22 (1983).
\vspace{-3.5 mm}
\item{[83P2]} J. Preskill (Monopoles in 1983
) Invited paper at the Monopole 83 Workshop (Ann Arbor, Mich., 1983).
\vspace{-3.5 mm}
\item{[83R1]} D. Ritson SLAC/PUB 2950/1983 (1983).
\vspace{-3.5 mm}
\item{[83R2]} Y. Rephaeli et al. (The magnetic monopole flux and the survival
of intracluster magnetic fields) Phys. Lett. B121 (1983) 115.
\vspace{-3.5 mm}
\item{[83R3]} T. W. Ruijgrok et al. (Monopole chemistry) Phys. Lett. B129
(1983) 209. 
\vspace{-3.5 mm}
\item{[83S1]} K. H. Schatten (Measurement of the magnetic monopole charge
of the moon) Phys. Rev. D27 (1983) 1525. 
\vspace{-3.5 mm}
\item{[83S2]} A. N. Schellekens et al. (Classical upper bounds for Grand
Unified monopole masses) Phys. Rev. Lett. 50 (1983) 1242.
\vspace{-3.5 mm}
\item{[83S3]} R. D. Sorkin (Kaluza-Klein monopole) Phys. Rev. Lett. 51 (1983)
87. 
\vspace{-3.5 mm}
\item{[83S4]} C. Schmid (Proton decay catalyzed by a monopole: the ratio of 
$e^+$ to $e^- \pi^0$ final states) Phys. Rev. D28 (1983) 1802. 
\vspace{-3.5 mm}
\item{[83T1]} G. Tiktopoulos (Atomic excitation and ionization by slow 
magnetic monopoles) Phys. Lett. B125 (1983) 156.
\vspace{-3.5 mm}
\item{[83T2]} C. D. Tesche et al. (Inductive monopole detector employing 
planar high order superconducting gradiometer coils) 
Appl. Phys. Lett. 43 (1983) 384.
\vspace{-3.5 mm}
\item{[83T3]} M. S. Turner (Monopole heat) Nature (London) 302 (1983) 804.
\vspace{-3.5 mm}
\item{[83V1]} M. A. Virasoro (On the S wave interaction between massive 
fermions and a GUT magnetic monopole) Phys. Lett. B125 (1983) 161. 
\vspace{-3.5 mm}
\item{[83Z1]} J. F. Ziegler et al. (Test of a superconducting magnetic monopole
detector for spurious events due to sea level cosmic rays) Phys. Rev. D28 (1983)
1793. 

\item{[84A1]} D. Akers (Magnetic monopole spin resonance) Phys. Rev. D29
(1984) 1026. 
\vspace{-3.5 mm}
\item{[84B1]} L. Bracci, G. Fiorentini (Interactions of magnetic monopoles
with nuclei and atoms: formation of bound states and phenomenological 
consequences) Nucl. Phys. B232 (1984) 236.
\vspace{-3.5 mm}

\item{[84B2]} L. Bracci et al. (On the energy loss of very slowly moving
magnetic monopoles) Nucl. Phys. B238 (1984) 167. 
\vspace{-3.5 mm}
\item{[84B3]} L. Bracci et al. (Formation of monopole - proton bound states
in the hot universe) Phys. Lett. B143 (1984) 357, erratum-ibid. B155 (1985)
468.
\vspace{-3.5 mm}
\item{[84B4]} A. P. Balachandran et al. (Monopole induced proton disintegration)
Phys. Rev. D29 (1984) 1184. 
\vspace{-3.5 mm}
\item{[84B5]} S. K. Bose (Selfdual monopoles in SU(5)) Phys. Rev. D30 (1984)
504. 
\vspace{-3.5 mm}
\item{[84B6]} C. Bernard et al. (Sonic search for monopoles, gravitational 
waves and newtorites) Nucl. Phys. B242 (1984) 93.
\vspace{-3.5 mm}
\item{[84C1]}  G. Callan et al. (Monopole catalysis of skyrmion decay) Nucl.
Phys. B239 (1984) 161.
\vspace{-3.5 mm}
\item{[84C2]} K. L. Chang et al. (Magnetic monopole and mixing angle in
Weinberg-Salam's theory) Lett. Nuovo Cim. 39 (1984) 23. 
\vspace{-3.5 mm}
\item{[84D1]} T. Datta (\v{C}erenkov magnon excitations by a subrelativistic
magnetic monopole) Phys. Lett. A103 (1984) 243.
\vspace{-3.5 mm}
\item{[84E1]} Z. F. Ezawa (Monopole - fermion dynamics and the Rubakov effect
in Kaluza-Klein theories) Phys. Lett. B138 (1984) 81.
\vspace{-3.5 mm}
\item{[84E2]} P. Eckert (The SU(5) monopole and its dependence on the weak 
scale) Nucl. Phys. B231 (1984) 40.
\vspace{-3.5 mm}
\item{[84F1]} G. Fiorentini (Molecular systems with muons or monopoles)
Nucl. Phys. A416 (1984). 
\vspace{-3.5 mm}
\item{[84F2]} J. N. Fry et al. (Supermassive monopole stars) Astrophys. J.
286 (1984) 397.
\vspace{-3.5 mm}
\item{[84G1]} G. Giacomelli (Magnetic monopoles) Nuovo Cim. 7 (1984) N.12,1.
\vspace{-3.5 mm}
\item{[84H1]} M. Honda (The potential between monopole and anti-monopole in
lattice space) Phys. Lett. B145 (1984) 243. 
\vspace{-3.5 mm}
\item{[84H2]} Z. Horvath et al. (SU(3) monopole of unit charge) Z. Phys. C22
(1984) 261. 
\vspace{-3.5 mm}
\item{[84I1]} J. Incandela et al. (Flux limit on cosmic ray magnetic monopoles
from a large area induction detector) Phys. Rev. Lett. 53 (1984) 2067. 
\vspace{-3.5 mm}
\item{[84K1]} M. R. Krishnaswamy et al. (Limits on the flux of monopoles
from the Kolar gold mine experiments) Phys. Lett. B142 (1984) 99.
\vspace{-3.5 mm}
\item{[84K2]} R. K. Kaul (Monopole mass in supersymmetric gauge theories)
Phys. Lett. B143 (1984) 427. 
\vspace{-3.5 mm}
\item{[84K3]} F. Kajino et al. (First results from a search for magnetic 
monopoles by a detector utilizing the Drell mechanism and the Penning effect)
Phys. Rev. Lett. 52 (1984) 1373.
\vspace{-3.5 mm}
\item{[84K4]} K. Kawagoe et al. (An underground search for anomalous slow
penetrating particles) Nuovo Cim. Lett. 41 (1984) 315. 
\vspace{-3.5 mm}
\item{[84L1]} T. M. Liss et al. (Unique search for Grand Unification 
magnetic monopoles) Phys. Rev. D30 (1984) 884. 
\vspace{-3.5 mm}
\item{[84L2]} G. Lazarides and Q. Shafi (Extended structures at intermediate 
scales in an inflationary cosmology) Phys. Lett. B148 (1984) 35.
\vspace{-3.5 mm}
\item{[84M1]} Z. Ma (Monopole and dyon solutions in SU(5) unified gauge
theory) Nucl. Phys. B231 (1984) 172.
\vspace{-3.5 mm}
\item{[84M2]} J. Madsen (Galaxy formation in a monopole dominated universe)
Phys. Lett. B143 (1984) 363. 
\vspace{-3.5 mm}
\item{[84N1]} A. J. Niemi et al. (On the electric charge of the magnetic
 monopole) Phys. Rev. Lett. 53 (1984) 515. 
\vspace{-3.5 mm}
\item{[84O1]} K. Olaussen et al. (Proton capture by magnetic monopoles) Phys.
Rev. Lett. 52 (1984) 325.
\vspace{-3.5 mm}
\item{[84O2]} P. Osland and T. T. Wu (Monopole-fermion and dyon-fermion 
bound states. 1. General properties and numerical results) Nucl. Phys.
B247 (1984) 421.
\vspace{-3.5 mm}
\item{[84O3]} P. Osland and T. T. Wu (Monopole-fermion and dyon-fermion
bound states. 2. Weakly bound states for the lowest angular momentum) Nucl. 
Phys. B247 (1984) 450.
\vspace{-3.5 mm}
\item{[84P1]} J. Preskill (Magnetic monopoles) Ann. Rev. Nucl. Part. Sci. 34
(1984) 461. 
\vspace{-3.5 mm}
\item{[84P2]} P. B. Price (Limit on flux of supermassive monopoles and charged
relic particles using plastic track detectors) Phys. Lett. B140 (1984) 112. 
\vspace{-3.5 mm}
\item{[84P3]} P. B. Price et al. (Search for GUT magnetic monopoles at a flux
level below the Parker limit) Phys. Rev. Lett. 52 (1984) 1265. 
\vspace{-3.5 mm}
\item{[84P4]} J. Polchinski (Monopole catalysis: the fermion rotor system)
Nucl. Phys. B242 (1984) 345. 
\vspace{-3.5 mm}
\item{[84R1]} G. A. Ringwood et al. (Monopoles admit parastatistics)
Phys. Rev. Lett. 53 (1984) 1980.
\vspace{-3.5 mm}
\item{[84R2]} V. A. Rubakov and M. S. Serebryakov (On the strong and weak
effects in the s-wave monopole-fermion interactions) Nucl. Phys. B237
(1984) 329.
\vspace{-3.5 mm}
\item{[84R3]} T. W. Ruijgrok (Binding of matter to a magnetic monopole)
Acta Phys. Polon. B15 (1984) 305.
\vspace{-3.5 mm}
\item{[84S1]} D. A. Sparrow et al. (Pion induced monopole transitions)
Phys. Rev. C29 (1984) 949. 
\vspace{-3.5 mm}
\item{[84S2]} A. N. Schellekens (Boundary-condition independence of catalysis
of proton decay by monopoles) Phys. Rev. D29 (1984) 2378.
\vspace{-3.5 mm}
\item{[84S3]} A. Sen (Role of conservation laws in the Callan-Rubakov  
processes with arbritary number of generations of fermions) 
Phys. Rev. Lett. 52 (1984) 1755.
\vspace{-3.5 mm}
\item{[84T1]} G. Tarle et al. (First results from a sea level search for
supermassive magnetic monopoles) Phys. Rev. Lett. 52 (1984) 90. 
\vspace{-3.5 mm}
\item{[84W1]} I. Wassermann et al. (Do monopoles cause rapid decay of the
 galactic field?) Cornell Univ. Reprint CRSR 810 (1984).
\vspace{-3.5 mm}
\item{[84W2]} T. F. Walsh et al. (Monopole catalysis of proton decay) Nucl.
Phys. B232 (1984) 349. 

\item{[85A1]} S. P. Ahlen et al. (Scintillation from slow protons: a probe
of monopole detection capabilities) Phys. Rev. Lett. 55 (1985) 181. 
\vspace{-3.5 mm}
\item{[85A2]} J. Arafune et al. (Monopole abundance in the Solar System and
the intrinsic heat in the Jovian planets) Phys. Rev. D32 (1985) 2586. 
\vspace{-3.5 mm}
\item{[85B1]} S. Bermon et al. (Flux limit of cosmic-ray magnetic monopoles
from a fully coincident superconductiviting induction detector) Phys. Rev.
Lett. 55 (1985) 1850.
\vspace{-3.5 mm}
\item{[85B2]} L. Bracci et al. (Monopole trapping inside stars and
 phenomenological consequences) Nucl. Phys. B258 (1985) 726.
\vspace{-3.5 mm}
\item{[85B3]}  L. Bracci and G. Fiorentini (Magnetic monopoles in stellar
interiors) Nuovo Cim. Lett. 42 (1985) 123.
\vspace{-3.5 mm}
\item{[85B4]} L. Bracci and G. Fiorentini (On the capture of protons by
magnetic monopoles) Nucl. Phys. B249 (1985) 519. 
\vspace{-3.5 mm}
\item{[85B5]} L. Bracci et al. (Magnetic monopoles and dyons interacting
with matter) Nucl. Phys. B259 (1985) 351. 
\vspace{-3.5 mm}
\item{[85B6]} L. Bracci et al. (Auger-like formation of monopolium) Phys.
Lett. B165 (1985) 425.
\vspace{-3.5 mm}
\item{[85B7]} S. K. Bose (Bound states of a charged particle and a dyon) 
J. Phys. A18 (1985) 1289.
\vspace{-3.5 mm}
\item{[85C1]} A. D. Caplin (New upper bound on flux of cosmic magnetic
monopoles) Nature 317 (1985) 234.
\vspace{-3.5 mm}
\item{[85C2]} G. Calucci and G. Vedovato (Pair production in the field
of a monopole with nonconserved quantum numbers) Z. Phys. C27 (1985) 377. 
\vspace{-3.5 mm}
\item{[85C3]} G. Calucci and G. Vedovato (Nucleon decay induced by GUT 
monopole and possible nonconservation of charge) Z. Phys. C29 (1985) 111. 
\vspace{-3.5 mm}
\item{[85C4]} B. Cabrera et al. (Sensing area ditribution functions for one
and three loop superconductive magnetic monopole detectors) Phys. Rev. D31
(1985) 2199. 
\vspace{-3.5 mm}
\item{[85E1]} T. Ebisu et al. (New limit on the magnetic monopole density
in old iron ore) J. Phys. G11 (1985) 883. 
\vspace{-3.5 mm}
\item{[85E2]} A. E. Everett et al. (Monopole annihilation and causality)
Phys. Rev. D31 (1985) 1925. 
\vspace{-3.5 mm}
\item{[85F1]} G. Fiorentini (Magnetic monopoles in ferromagnetic materials)
Nucl. Phys. B262 (1985) 49. \par
\vspace{-3.5 mm}
\item{[85F2]} J. Fineberg (Monopole pair production in compact U(1)) Phys.
Lett. B158 (1985) 135. 
\vspace{-3.5 mm}
\item{[85I1]} J. Incandela et al. (First results from a 1.1 m diameter
superconducting monopole detector) EFI 85-75 (1985). 
\vspace{-3.5 mm}
\item{[85K1]} C. W. Kim et al. (Avoiding the magnetic monopole problem in the
new inflation theories: the 75 of SU(5)) Phys. Lett. B163 (1985) 87. 
\vspace{-3.5 mm}
\item{[85K2]} F. Kajino and Y. K. Yuan (Proportional chambers utilizing the
Drell mechanism and the Penning effect to search for GUT monopoles) Nucl.
Instr. Meth. A228 (1985) 278.
\vspace{-3.5 mm}
\item{[85M1]} Z. Ma et al. (On the catalysis effect of magnetic monopole for 
proton decay) Phys. Lett. B153 (1985) 59.
\vspace{-3.5 mm}
\item{[85M2]} N. S. Manton (Monopole interactions at long range) Phys. Lett.
B154 (1985) 397, erratum-ibid B157 (1985) 475. 
\vspace{-3.5 mm}
\item{[85O1]} P. Osland et al. (Monopole-fermion and dyon-fermion bound states.
6. Weakly bound states for the dyon-fermion system) Nucl. Phys. B261 (1985)
687.  
\vspace{-3.5 mm}
\item{[85S1]} A. Sen (Monopole-induced baryon number violation due to weak
anomaly) Nucl. Phys. B250 (1985) 1.
\vspace{-3.5 mm}
\item{[85S2]} J. L. Stone et al. (Experimental limits on monopole catalysis,
$N-\bar{N}$ oscillations and nucleon lifetimes) Nucl. Phys. B252 (1985) 261. 
\vspace{-3.5 mm}
\item{[85S3]} P. Salomonson et al. (On the primordial monopole problem in
Grand Unified Theories) Phys. Lett. B151 (1985) 243. 
\vspace{-3.5 mm}
\item{[85T1]} J. F. Tang (Magnetic monopole and its catalysis effect on
baryon number nonconservation process) Commun. Theor. Phys. 4 (1985) 631.
\vspace{-3.5 mm}
\item{[85W1]} G. Wanders (The state space of the fermion - monopole system)
Nucl. Phys. B255 (1985) 174.

\item{[86A1]} M. Aglietta et al. (Monopole search with the Mont Blanc
LSD experiment) Nuovo Cim. 9C (1986) 588.
\vspace{-3.5 mm}
\item{[86A2]} D. Akers (Magnetic monopole interactions: shell structure of
meson and baryon states) Int. J. Theor. Phys. 25 (1986) 1281. 
\vspace{-3.5 mm}
\item{[86B1]} G. Battistoni et al. (Limit on monopole flux in the Mont Blanc
NUSEX experiment) Nuovo Cim. 9C (1986) 551. 
\vspace{-3.5 mm}
\item{[86B2]} S. K. Bose (Dyon electron bound states
) J. Phys. G12 (1986) 1135.
\vspace{-3.5 mm}
\item{[86C1]} A. D. Caplin et al. (Observation of an unexplained event from
a magnetic monopole detector - another monopole candidate?) Nature 321 
(1986) 402. 
\vspace{-3.5 mm}
\item{[86C2]} N. S. Craigie, G. Giacomelli, W. Nahm and Q. Shafi 
(Theory and detection of magnetic monopoles
in gauge theories) World Scientific (1986) 499. 
\vspace{-3.5 mm}
\item{[86C3]} M. W. Cromar et al. (Flux limit of cosmic-ray magnetic monopoles
from a multiply discriminating superconducting detectors) Phys. Rev. Lett.
56 (1986) 2561. 
\vspace{-3.5 mm}
\item{[86D1]} A. M. Defaria-Rosa et al. (A satisfactory formalism for
magnetic monopoles by Clifford algebras) Phys. Lett. B173 (1986) 233.
\vspace{-3.5 mm}
\item{[86G1]} M. Goldhaber et al. (Grand Unification, proton decay and magnetic 
monopoles) Comments Nucl. Part. Phys. 16 (1986) 23.  
\vspace{-3.5 mm}
\item{[86G2]} D. E. Groom (In search of the supermassive magnetic monopole)
Phys. Rep. 140 (1986) 323.
\vspace{-3.5 mm}
\item{[86H1]} T. Hara et al. (Slow-monopole search with large-area helium-gas
proportional-counter array) Phys. Rev. Lett. 56 (1986) 553. 
\vspace{-3.5 mm}
\item{[86H2]} M. Henneaux et al. (Quantization of topological mass in the
presence of a magnetic pole) Phys. Rev. Lett. 56 (1986) 689. 
\vspace{-3.5 mm}
\item{[86H3]} Huan-Hua Cui et al. (The search for magnetic monopoles in
magnetite from North China) Chin. Phys. Lett. 3 (1986) 461. 
\vspace{-3.5 mm}
\item{[86H4]} J. A. Harvey et al. (Effects of neutron-star superconductivity on
magnetic monopoles and core field decay) Phys. Rev. D33 (1986) 2084.
\vspace{-3.5 mm}
\item{[86I1]} J. Incandela et al. (First results from a 1.1-m-diameter
superconducting monopole detector) Phys. Rev. D34 (1986) 2637. 
\vspace{-3.5 mm}
\item{[86I2]} M. Izawa (On stars which burn by the Rubakov process) Progr.
Theor. Phys. 75 (1986) 556. \par 
\vspace{-3.5 mm}
\item{[86K1]} T. W. B. Kibble (String-dominated universe) Phys. Rev. D33
(1986) 328.
\vspace{-3.5 mm}
\item{[86L1]} G. Lazarides et al. (Magnetic monopoles from superstring models)
Bartol. Res. Found. 2710 (1986) 11. 
\vspace{-3.5 mm}
\item{[86L2]} H. J. Lipkin et al. (Magnetic monopoles, electric currents and
Dirac strings) Phys. Lett. B179 (1986) 13. 
\vspace{-3.5 mm}
\item{[86L3]} D. London (Is the doubly charged monopole stable?) Phys. Rev.
D33 (1986) 3075. 
\vspace{-3.5 mm}
\item{[86L4]} X. Li et al. (Generalized y dependent monopole in (4+k)-
dimension abelian theories) Phys. Rev. D34 (1986) 1124. 
\vspace{-3.5 mm}
\item{[86M1]} L. G. Mestres et al. (Detection of magnetic monopoles with
superheated type-I superconductors) Nuovo Cim. 9C (1986) 573.
\vspace{-3.5 mm}
\item{[86M2]} P. Musset (Magnetic monopoles) Nuovo Cim. 9C (1986) 559.
\vspace{-3.5 mm}
\item{[86P1]} P. B. Price et al. (Search for supermassive magnetic monopoles
using mica crystals) Phys. Rev. Lett. 56 (1986) 1226.
\vspace{-3.5 mm}
\item{[86R1]} L. E. Roberts (Dirac magnetic monopole pair production in
relativistic nucleus nucleus collisions) Nuovo Cim. 92A (1986) 247.
\vspace{-3.5 mm}
\item{[86S1]} J. Sanudo et al. (Magnetic monopoles and strange matter) Phys.
Lett. B166 (1986) 45. 
\vspace{-3.5 mm}
\item{[86S2]} R. D. Sorkin (Topology charge and monopole creation) Phys. Rev.
D33 (1986) 978. 
\vspace{-3.5 mm}
\item{[86S3]} M. K. Sundaresan et al. (Kaluza-Klein monopoles in five
 dimensions) Phys. Rev. D33 (1986) 484.
\vspace{-3.5 mm}
\item{[86S4]} J. Seixas (Magnetic monopoles: from classical to quantum physics)
Phys. Lett. B171 (1986) 95. 
\vspace{-3.5 mm}
\item{[86T1]} C. Teitelboim (Monopoles of higher rank) Phys. Lett. B167
(1986) 69.
\vspace{-3.5 mm}
\item{[86T2]} S. Tasaka and T. Suda (Deep underground search for massive
magnetic monopoles by CR-39 plastic track detector) Jour. Phys. Soc. Jap.
55 (1986) 3749.
\vspace{-3.5 mm}
\item{[86U1]} W. G. Unruh (Accelerated monopole detector in odd space-time
dimensions) Phys. Rev. D34 (1986) 1222. 

\item{[87A1]} D. Akers (Further evidence for magnetic charge from meson
spectroscopy) Int. J. Theor. Phys. 26 (1987) 1169. 
\vspace{-3.5 mm}
\item{[87B1]} B. Barish et al. (Search for Grand Unification monopoles
and other ionizing heavy particles using a scintillation detector at the 
Earth's surface) Phys. Rev. D36 (1987) 2641. 
\vspace{-3.5 mm}
\item{[87B2]} J. Bartelt et al. (Monopole-flux and proton-decay limits from
the Soudan 1 detector) Phys. Rev. D36 (1987) 1990. 
\vspace{-3.5 mm}
\item{[87C1]} E. Comay (Geometry and charge-monopole systems) Phys. Lett.
B187 (1987) 111. 
\vspace{-3.5 mm}
\item{[87E1]} T. Ebisu et al. (Search for magnetic monopoles trapped in old
iron ores using a superconducting detector) Phys. Rev. D36 (1987) 3359.
\vspace{-3.5 mm}
\item{[87F1]} D. J. Ficenec (Observation of electronic exitation by extremely
slow protons with applications to the detection of supermassive charged
particles) Phys. Rev. D36 (1987) 311. 
\vspace{-3.5 mm}
\item{[87F2]} S. Fitzsimmons et al. (Properties of some selfdual monopoles)
Phys. Rev. D36 (1987) 2571. \par 
\vspace{-3.5 mm}
\item{[87G1]} H. B. Gao et al. (Topological quantization of mass in extended
gauge theories with a monopole) Phys. Lett. B188 (1987) 105. 
\vspace{-3.5 mm}
\item{[87G2]} T. Gentile et al. (Search for magnetically charged particles
produced in $e^+e^-$ annihilations at $\sqrt s=10.6$ GeV) Phys. Rev. D35 (1987)
1081. 
\vspace{-3.5 mm}
\item{[87H1]} H. M. Hodges et al. (Parker limit for monopoles with large 
magnetic charge) Phys. Rev. D35 (1987) 2024.
\vspace{-3.5 mm}
\item{[87I1]} K. Isler et al. (Monopole core excitations and the
 Rubakov-Callan effect) Nucl. Phys. B294 (1987) 925.
\vspace{-3.5 mm}
\item{[87K1]} A. S. Kronfeld et al. (Monopole condensation and color
confinement) Phys. Lett. B198 (1987) 516. 
\vspace{-3.5 mm}
\item{[87L1]} G. Lazarides et al. (Magnetic monopoles from superstring
models) Phys. Rev. Lett. 58 (1987) 1707.
\vspace{-3.5 mm}
\item{[87M1]} R. Montgomery (Correction to the low-energy scattering of
 monopoles) Phys. Lett. A125 (1987) 159. 
\vspace{-3.5 mm}
\item{[87M2]} N. S. Manton (Monopole and skyrmion bound states) Phys. Lett.
B198 (1987) 226. 
\vspace{-3.5 mm}
\item{[87M3]} G. E. Masek et al. (Results from a magnetic monopole search
utilizing helium proportional tubes) Phys. Rev. D35 (1987) 2758. 
\vspace{-3.5 mm}
\item{[87N1]} S. Nakamura et al. (Search for supermassive relics) Phys. Lett.
B183 (1987) 395.
\vspace{-3.5 mm}
\item{[87N2]} G. Ni and Y. Cen (On the electric charge of the Dirac dyon)
Phys. Lett. B188 (1987) 236.
\vspace{-3.5 mm}
\item{[87O1]} M. Oleszczuk et al. (Monopole-antimonopole pair solution of a
classical SU(3) Yang-Mills theory) Phys. Lett. D35 (1987) 3225. 
\vspace{-3.5 mm}
\item{[87P1]} E. Papageorgiu et al. (Identification of magnetic monopoles
via electron-positron pair production) Phys. Lett. B197 (1987) 277.
\vspace{-3.5 mm}
\item{[87P2]} P. B. Price et al. (Search for highly ionizing particles at
the Fermilab proton-antiproton collider) Phys. Rev. Lett. 59 (1987) 2523.  
\vspace{-3.5 mm}
\item{[87S1]} J. C. Schouten et al. (Design and performance of a 0.18 $m^2$
inductive detector for cosmic magnetic monopoles) J. Phys. E20 (1987) 850. 
\vspace{-3.5 mm}
\item{[87S2]} M. J. Shepko et al. (Search for superheavy Grand Unified 
magnetic monopoles in cosmic rays) Phys. Rev. D35 (1987) 2917.
\vspace{-3.5 mm}
\item{[87T1]} T. Tsukamoto et al. (Limits on the flux of supermassive relics)
Europhys. Lett. 3 (1987) 39.
\vspace{-3.5 mm}
\item{[87Z1]} H. Zhang (On the $\theta$ term effect on the charge spectra of
dyons associated with generalized magnetic monopoles and on quarks as dyons
in a spontaneously broken gauge theory) Phys. Rev. D36 (1987) 1868. 

\item{[88A1]} D. Akers (Mikhailov's experiments on detection of magnetic charge)
Int. J. Theor. Phys. 27 (1988) 1019. 
\vspace{-3.5 mm}
\item{[88A2]} G. Auriemma et al. (Monopole trigger for the streamer tube system
in MACRO) Nucl. Instr. Meth. A263 (1988) 249.
\vspace{-3.5 mm}
\item{[88B1]} Blagojevic et al. (The quantum field theory of electric
and magnetic charge) Phys. Rep. 157 (1988) 233. 
\vspace{-3.5 mm}
\item{[88B2]} W. Braunschweig et al. (A search for particles with magnetic 
charge produced in $e^+e^-$ annihilations at $\sqrt s=35$ GeV) Z. Phys.
C38 (1988) 543. 
\vspace{-3.5 mm}
\item{[88B3]} E. Bellotti (The Gran Sasso underground laboratory) 
Nucl. Instr. Meth. A264 (1988) 1.
\vspace{-3.5 mm}
\item{[88B4]} G. Battistoni et al. (Response of streamer tubes to highly
ionizing particles) Nucl. Instr. Meth. A270 (1988) 185.
\vspace{-3.5 mm}
\item{[88C1]} E. Copeland et al. (Monopoles connected by strings and the
 monopole problem) Nucl. Phys. B298 (1988) 445.
\vspace{-3.5 mm}
\item{[88D1]} T. Doke et al. (CR-39 plastic for massive magnetic monopole
search) Nucl. Instr. Meth. Res. B34 (1988) 81.
\vspace{-3.5 mm}
\item{[88G1]} A. S. Goldhaber et al. (Is monopole catalyzed nucleon decay
detectable?) Nucl. Phys. B296 (1988) 955. 
\vspace{-3.5 mm}
\item{[88K1]} K. Kinoshita et al. (Search for highly ionizing particles in 
$e^+e^-$
annihilations at $\sqrt s=50-52$ GeV) Phys. Rev. Lett. 60 (1988) 1610.
\vspace{-3.5 mm}
\item{[88L1]} J. P. M. Lebrun (Electrodynamics of magnetic charge 
distributions for applications to chiral lagrangian theory and electromagnetic
properties of hadrons) Nuovo Cim. 99A (1988) 211.  
\vspace{-3.5 mm}
\item{[88M1]} H. Mishra et al. (Quantum magnetic monopoles) Ind. J. Phys.
62A (1988) 420.
\vspace{-3.5 mm}
\item{[88O1]} K. Ogura et al. (Improvements of CR-39 for massive magnetic 
monopoles search) Nucl. Tracks Radiat. Meas. 15 (1988) 315.
\vspace{-3.5 mm}
\item{[88S1]} K. S. Somalwar et al. (Transient response induction detectors
for magnetic monopoles: first operation at 78 $^0K$) Phys. Rev. D  (1988) 2403. 

\item{[89A1]} L. W. Alvarez et al. (Search for magnetic monopoles in the lunar
sample) Trower, W. P. : Discovering Alvarez (1989) 178.
\vspace{-3.5 mm}
\item{[89B1]} Y. Benadjal (Search for magnetic monopoles with the Frejus
detector) LAL-89-69, Oct. 1989, Doctoral Thesis. 
\vspace{-3.5 mm}
\item{[89B2]} J. Bartelt (Monopole flux and proton decay limits from Soudan-I
detector) Phys. Rev. D40 (1989) 1701. 
\vspace{-3.5 mm}
\item{[89C1]} B. Cabrera et al. (A superconductive detector to search
for cosmic ray magnetic monopoles) Fairbank, J. D. et al. : Near zero:
New frontiers of physics (1989) 546.  
\vspace{-3.5 mm}
\item{[89D1]} T. Doke et al. (Search for GUTs monopoles with track etch
 detectors) Nucl. Tracks Radiat. Meas. 16 (1989) 107.
\vspace{-3.5 mm}
\item{[89F1]} Fujii et al. (Search for highly ionizing particles
in e$^+$e$^-$ annihilations at $S^{1/2}=50$  GeV to 60.8 GeV) 
 Phys. Lett. B228 (1989) 543. 
\vspace{-3.5 mm}
\item{[89F2]} M. A. Faria-Rosa and W. A. Rodrigues (A geometrical theory
of nontopological magnetic monopoles) Mod. Phys. Lett. A4 (1989) 175. 
\vspace{-3.5 mm}
\item{[89F3]} L. Feher et al. (Separating the dyon system) Phys. Rev.
D40 (1989) 666. 
\vspace{-3.5 mm}
\item{[89L1]} J. P. M. Lebrun (Differentials of magnetic charge currents
and consequent revision of electric charge quantization rule) Nuovo
Cim. 102A (1989) 925.
\vspace{-3.5 mm}
\item{[89M1]} A. M. R. Magnon (The magnetic charge phenomenon: comparison
of theory and experiment) Nuovo Cim. 102A (1989) 964. 
\vspace{-3.5 mm}
\item{[89O1]} K. Ogura et al. (Search of GUTs monopoles with track etch 
detectors) Radiat. Meas. (1989) 107.  
\vspace{-3.5 mm}
\item{[89P1]}  V. Patera (Excitation and ionization in low-Z atoms
by slow magnetic monopoles) Phys. Lett. 137A (1989) 259.

\item{[90A1]} S. P. Ahlen et al. (Results from the MACRO detector at the 
Gran Sasso Laboratory) Nucl. Phys. B16 (1990) 486. 
\vspace{-3.5 mm}
\item{[90A2]} E. N. Alexeyev et al. (A search for superheavy magnetic monopole
by Baksan underground scintillation telescope) ICRC, Adelaide, vol. 10
 (1990) 83.
\vspace{-3.5 mm}
\item{[90A3]}  H. Adarkar et al. (Kolar gold field monopole experiment) 21st
 ICRC Conf. Papers 10 (1990) 95.
\vspace{-3.5 mm}
\item{[90A4]} D. Akers (Detection of the Dirac monopole with magnetic
 levitation) Int. J. Theor. Phys. 29 (1990) 109.
\vspace{-3.5 mm}
\item{[90A5]} D. Akers (Existence of magnetic charge) Int. J. Theor.
 Phys. 29 (1990) 1091. 
\vspace{-3.5 mm}
\item{[90A6]} O. Abe (Response function of accelerated monopole detector
in R $\times$ T(3) space-time) Phys. Rev. D41 (1990) 1897.
\vspace{-3.5 mm}
\item{[90B1]} S. Bermon et al. (New limit set on cosmic-ray monopole
flux by a large-area superconducting magnetic-induction detector) Phys. Rev.
 Lett. 64 (1990) 839.
\vspace{-3.5 mm}
\item{[90B2]} K. N. Buckland et al. (Results of a magnetic-monopole search
utilizing a large-area proportional-tube array) Phys. Rev. D41 (1990) 2726.
 \vspace{-3.5 mm}
\item{[90B3]} Bertani et al. (Search for magnetic monopoles at the Tevatron
Collider) Europhys. Lett. 12 (1990) 613.
\vspace{-3.5 mm}
\item{[90B4]} L. B. Bezrukov et al. (Search for superheavy magnetic monopoles
in deep underwater experiments at lake Baikal) Yad. Fiz. 52 (1990) 86.
\vspace{-3.5 mm}
\item{[90B5]} S. K. Bose and C. C. Choo (The dyon - electron system: 
scattering and electron capture) J. Phys. A23 (1990) 2961. 
\vspace{-3.5 mm}
\item{[90C1]} B. Cabrera et al. (Limit on the flux of cosmic ray magnetic
monopoles from operation of an eight loop superconducting detector) Phys. 
Rev. Lett. 64 (1990) 835.
\vspace{-3.5 mm}
\item{[90C2]} M. Calicchio et al. (Status report of the MACRO experiment at
Gran Sasso) Nucl. Phys. Proc. Suppl. 13 (1990) 368. 
\vspace{-3.5 mm}
\item{[90C3]} E. Comay (Charges, monopoles and unit systems) Acta Phys. 
Polon. B21 (1990) 171.
\vspace{-3.5 mm}
\item{[90D1]} T. Doke et al. (CR-39 detector and experimental techniques of
cosmic supermassive particles search) Nucl. Instr. Meth. A286 (1990) 327.
\vspace{-3.5 mm}
\item{[90D2]} A. Davidson and E. Guendelman (Electric monopole with
internal magnetic monopole like structure) Phys. Lett. B251 (1990) 250.
\vspace{-3.5 mm}
\item{[90E1]} A. A. Ershov and D. V. Galtsov (Nonexistence of regular
monopoles and dyons in the SU(2) Einstein Yang-Mills theory) Phys. Lett.
A150 (1990) 159.
\vspace{-3.5 mm}
\item{[90F1]} D. J. Ficenec (A search for magnetic monopoles and other
supermassive charged particles) UMI-90-23762-mc Ph.D. Thesis, Boston 
University Graduate School (1990).
\vspace{-3.5 mm}
\item{[90G1]} D. Ghosh and S. Chatterjea (Supermassive magnetic monopoles
flux from the oldest mica samples) Europhys. Lett. 12 (1990) 25.
 \vspace{-3.5 mm}
\item{[90H1]} T. Hara et al. (A search experiment for slow moving 
monopoles ($\beta\geq 2\times 10^{-4})$ using helium proportional counters
 array) 21st ICRC Conf. Papers 10 (1990) 79.
\vspace{-3.5 mm}
\item{[90I1]} H. Ichinose et al. (CR-39 detector and experimental techniques 
of cosmic supermassive particles search) Nucl. Instrum. Meth. A286 (1990) 327.
\vspace{-3.5 mm}
\item{[90K1]} V. K. Korshunov (Drift motion of the magnetic monopole of
Poljakov-'t Hooft in the air and the ``ball-lightning" phenomenon) Mod.
Phys. Lett. A21, Vol. 5 (1990) 1629.  
\vspace{-3.5 mm}
\item{[90K2]} D. A. Kirzhnits (Slowing of a magnetic monopole in matter)
Sov. Phys. JETP 71 (1990) 427.
\vspace{-3.5 mm}
\item{[90K3]} V. G. Kiselev (A monopole in the Coleman-Weinberg model)
Phys. Lett. B249 (1990) 269. 
\vspace{-3.5 mm}
\item{[90L1]} N. P. Landsman (Quantization and superselection sectors.
2. Dirac monopole and Aharonov-Bohm effect) Rev. Math. Phys. 2 (1990) 73. 
\vspace{-3.5 mm}
\item{[90L2]} J. Q. Liang and D. S. Kulshreshtha (A charged particle
in the magnetic field of a Dirac monopole line) Phys. Lett. A149 (1990)
1.  
\vspace{-3.5 mm}
\item{[90M1]} A. M. Matheson and D. M. Upton (Monopole accretion
by cosmic strings) Mod. Phys. Lett. A5 (1990) 1313. 
\vspace{-3.5 mm}
\item{[90O1]} H. A. Olsen et al. (On the existence of bound states for a
massive spin 1 particle and a magnetic monopole) Phys. Rev. D42 (1990)
665. 
\vspace{-3.5 mm}
\item{[90O2]} H. A. Olsen and P. Osland (Bound states for a massive spin
1 particle and a magnetic monopole) Phys. Rev. D42 (1990) 690. 
\vspace{-3.5 mm}
\item{[90O3]} K. A. Olive (Inflation) Phys. Rep. 190 (1990) 307.
\vspace{-3.5 mm}
\item{[90S1]} M. Spurio (Ricerca di monopoli magnetici nell' esperimento
MACRO al Gran Sasso) Ph.D. Thesis, University of Bologna (1990).

\item{[91A1]}  S. P. Ahlen et al. (Improvements in the CR39 polymer for the 
MACRO experiment at the Gran Sasso Laboratory)
  Nucl. Tracks Radiat.
 Meas. 19 (1991) 641.
 \vspace{-3.5 mm}
\item{[91B1]} R. Bellotti et al. (First results from the MACRO experiment at
the Gran Sasso) Nucl. Phys. (Proc. Suppl.) 19 (1991) 128. 
\vspace{-3.5 mm}
\item{[91B2]} N. A. Batakis and A. A. Kehagias (A class of monopole vacua)
Phys. Lett. B271 (1991) 68. \par 
\vspace{-3.5 mm}
\item{[91B3]} S. K. Bose (Cosmological formation of dyon fermion bound states)
J. Phys. A24 (1991) 3711. \par 
\vspace{-3.5 mm}
\item{[91B4]} L. Brekke et al. (Alice strings, magnetic monopoles and charge
quantization) Phys. Rev. Lett. 67 (1991) 3643. 
\vspace{-3.5 mm}
\item{[91C1]} B. Cabrera et al. (Search for cosmic ray magnetic monopoles
using a three loop superconductive detector) Phys. Rev. D44 (1991) 622. 
\vspace{-3.5 mm}
\item{[91C2]} B. Cabrera et al. (Search for cosmic ray magnetic monopoles
with an eight channel superconducting detector) Phys. Rev. D44 (1991) 636. 
\vspace{-3.5 mm}
\item{[91C3]} H. C. Chandola et al. (Supersymmetric dyons) Phys. Rev.
D43 (1991) 3550. 
\vspace{-3.5 mm}
\item{[91G1]} H. Gao et al. (Detector for magnetic monopoles at OPAL)
Nucl. Instr. Meth. A302 (1991) 434. 
\vspace{-3.5 mm}
\item{[91G2]} J. M. Getino et al. (Interaction between electric currents
and magnetic monopoles) Europhys. Lett. 15 (1991) 821. 
\vspace{-3.5 mm}
\item{[91G3]} S. Graf et al. (Mass limit for Dirac type magnetic monopoles)
Phys. Lett. B262 (1991) 463. \par 
\vspace{-3.5 mm}
\item{[91H1]} S. J. Hands et al. (Monopole condensation and the dynamics
of chiral symmetry breaking in noncompact lattice QED with dynamical
fermions) Phys. Lett. B261 (1991) 294. 
\vspace{-3.5 mm}
\item{[91K1]} T. W. B. Kibble and E. J. Weinberg (When does causality
constrain the monopole abundance?) Phys. Rev. D43 (1991) 3188. 
\vspace{-3.5 mm}
\item{[91N1]} S. Nakariki (A monopole solution in Poincare gauge theory)
Nuovo Cim. 106B (1991) 945. 
\vspace{-3.5 mm}
\item{[91O1]} S. Orito et al. (Search for supermassive relics with
a 2000-$m^2$ array of plastic track detectors) Phys. Rev. Lett. 66 (1991)
 1951.
\vspace{-3.5 mm}
\item{[91P1]} L. Patrizii et al. (Improvements for the CR39 polymer for the 
MACRO experiment at the Gran Sasso laboratory) Radiat. Meas. 19 (1991) 641.

\item{[92A1]}  M. Ambrosio et al. (The QTP system for the MACRO experiment
 at Gran Sasso) Nucl. Instr. Meth. A321 (1992) 609.
\vspace{-3.5 mm}
\item{[92A2]} D. Akers (Dirac monopole and Mac Gregor's formula for particle
lifetimes) Nuovo Cim. 105A (1992) 935.  
\vspace{-3.5 mm}
\item{[92A3]} S. Ahlen et al., MACRO Coll.
 (Search for nuclearites using the MACRO detector) Phys. Rev. Lett.
69 (1992) 1860.  
\vspace{-3.5 mm}
\item{[92B1]} J. Baacke (Fluctuations and stability of the 't Hooft-Polyakov
monopole) Z. Phys. C53 (1992) 399. 
\vspace{-3.5 mm}
\item{[92C1]} A. Casher and Y. Shamir (Supersymmetry violation in elementary
particle - monopole scattering) Phys. Lett. B274 (1992) 381. 
\vspace{-3.5 mm}
\item{[92C2]} P. Chomaz et al. (From experimental monopole cross-sections
to nuclear incompressibilities) Phys. Lett. B281 (1992) 6. 
\vspace{-3.5 mm}
\item{[92C3]} H. C. Chandola et al. (Bound state of pointlike dyons)
Indian J. Pure Appl. Phys. 30 (1992) 193. 
\vspace{-3.5 mm}
\item{[92D1]} V. Dixit and M. Sher (Monopole annihilation and baryogenesis
at the electroweak scale) Phys. Rev. Lett. 68 (1992) 560. 
\vspace{-3.5 mm}
\item{[92D2]} A. C. Davis et al. (Monopole baryogenesis in the langacker-pi
scenario) Phys. Lett. B293 (1992) 123. 
\vspace{-3.5 mm}
\item{[92F1]} T. H. Farris et al. (The minimal electroweak model for monopole
annihilation) Phys. Rev. Lett. 68 (1992) 564.
\vspace{-3.5 mm}
\item{[92G1]} E. Gates et al. (Monopole annihilation at the electroweak
scale) Phys. Lett. B284 (1992) 309. 
\vspace{-3.5 mm}
\item{[92H1]} R. Holman et al. (How efficient is the langacker-pi
mechanism of monopole annihilation?) Phys. Rev. Lett. 69 (1992) 241. 
\vspace{-3.5 mm}
\item{[92K1]} H. Kataoka et al. (A dynamics of monopole in gauge space
and quantization of dyon charge) Mod. Phys. Lett. A7 (1992) 2165.  
\vspace{-3.5 mm}
\item{[92P1]} Particle Data Group (Review of particle physics)
Phys. Rev. D45 (1992) 14.
\vspace{-3.5 mm}
\item{[92S1]} D. P. Snowden-Ifft and P. B. Price (The low velocity response
of the solid state nuclear track detector CR39) Phys. Lett. B288 (1992) 250.
\vspace{-3.5 mm}
\item{[92S2]} Y. M. Shnir et al. (P violating magnetic monopole influence
on the behavior of the atom like system in external fields) Int. J. Mod.
Phys. A7 (1992) 3747. 
\vspace{-3.5 mm}
\item{[92T1]} J. L. Thron et al. (A search for magnetic monopoles with
the SOUDAN-2 detector) Phys. Rev. D46 (1992) 4846.

\item{[93A1]} S. Ahlen et al. (First supermodule of the MACRO
detector at Gran Sasso) Nucl. Instr. Meth. in Phys. Res. A324
 (1993) 337.
\vspace{-3.5 mm}
\item{[93A2]} F. C. Adams et al. (Extension of the Parker bound on the 
flux of magnetic monopoles) Phys. Rev. Lett. 70 (1993) 2511.
\vspace{-3.5 mm}
\item{[93B1]} A. O. Barut et al. (The Lamb shift in the charge magnetic
monopole system) Mod. Phys. Lett. A8 (1993) 3443. 
\vspace{-3.5 mm}
\item{[93B2]} D. Bak and C. Lee (Scattering of light by a BPS monopole)
Nucl. Phys. B403 (1993) 315. 
\vspace{-3.5 mm}
\item{[93D1]} T. Dobbins et al. (Update estimate of limits on the production 
rates of 't Hooft-Polyakov magnetic monopoles) Nuovo Cim. 106A (1993) 1295.
\vspace{-3.5 mm}
\item{[93H1]} A. Herdegen (Angular momentum in electrodynamics and an
argument against the existence of magnetic monopoles) J. Phys. A26
(1993) L449. 
\vspace{-3.5 mm}
\item{[93H2]} J. Hong (Search for GUT magnetic monopoles and other 
supermassive particles with the MACRO detector) Ph.D. Thesis, Cal. Institute
of Technology (1993).
\vspace{-3.5 mm}
\item{[93K1]} C. A. Kocher (Magnetic monopoles and hyperfine structure)
Ann. J. Phys. 61 (1993) 879.
\vspace{-3.5 mm}
\item{[93L1]} Yu M. Loskutov (Evolution of an homogeneus isotropic universe,
dark matter, and the absence of monopoles) Theor. Math. Phys. 94 (1993) 358.
\vspace{-3.5 mm}
\item{[93S1]} A. Sen (Quantization of dyon charge and electric magnetic
duality in string theory) Phys. Lett. B303 (1993) 22. 
\vspace{-3.5 mm}
\item{[93S2]} V. Singh (Magnetic monopoles) Indian J. Phys. 67 (1993) 45.
\vspace{-3.5 mm}
\item{[93S3]} R. Singer and D. Trautmann (Excitation of atoms by slow magnetic 
monopoles: a fully numerical Hatree-Fock approach) Nucl. Phys. A554 (1993) 421.

\item{[94A1]} S. Ahlen et al. (Search for slowly moving monopoles with the MACRO
 detector) Phys. Rev. Lett. 72 (1994) 608.
\vspace{-3.5 mm}
\item{[94A2]} I. Affleck and J. Sagi (Monopole catalyzed baryon decay: a
boundary conformal field approach) Nucl. Phys. B417 (1994) 374.
\vspace{-3.5 mm}
\item{[94B1]} R. Becker-Szendy et al. (New magnetic monopole flux limits from
 the IMB proton decay detector) Phys. Rev. D49 (1994) 2169.  
\vspace{-3.5 mm}
\item{[94B2]} K. Behrndt (A monopole solution in open string theory)
Nucl. Phys. B414 (1994) 114. 
\vspace{-3.5 mm}
\item{[94B3]} A. D. Barut et al. (Enhancement of the rate of radiative 
processes in the field of a magnetic monopole) Phys. Scripta 49 (1994) 513.
\vspace{-3.5 mm}
\item{[94B4]} B. C. Barish et al. (Reply to ``Flux limits for supermassive  
magnetic monopoles") Phys. Rev. Lett. 73 (1994) 1306.
\vspace{-3.5 mm}
\item{[94D1]} S. Das and P. Majundar (Charge-monopole versus gravitational
scattering at Planckian energies) Phys. Rev. Lett. 72 (1994) 2524.
\vspace{-3.5 mm}
\item{[94G1]} G. Giacomelli (Magnetic monopole searches)
Lectures at the 1994 Lake Louise Winter Institute, DFUB 20/94 (1994).
\vspace{-3.5 mm}
\item{[94H1]} S. J. Hands et al. (Spectroscopy, equation of state and
monopole percolation in lattice QED with two flavors) Nucl. Phys. B413
(1994) 503. 
\vspace{-3.5 mm}
\item{[94L1]} A. Linde (Monopoles as big as a universe) Phys. Lett. B327
(1994) 208.
\vspace{-3.5 mm}
\item{[94M1]} E. C. Marino and R. O. Ramos (Mass spectrum and correlation 
function of non abelian quantum magnetic monopoles) Phys. Rev. D49 (1994) 1093.
\vspace{-3.5 mm}
\item{[94M2]} A. Maia and W. A. Rodrigues (A generalized Dirac's quantization 
condition for phenomenological nonabelian magnetic monopoles) 
Mod. Phys. Lett. A9 (1994) 81.
\vspace{-3.5 mm}
\item{[94P1]} P. B. Price (Flux limits for supermassive magnetic monopoles)
Phys. Rev. Lett. 73 (1994) 1305.
\vspace{-3.5 mm}
\item{[94P2]} Particle Data Group (Review of particle physics)
Phys. Rev. D50 (1994) 1251.
\vspace{-3.5 mm}
\item{[94R1]} H. Ren (Fermions in a global monopole background) Phys. Lett.
 B325 (1994) 149. 

\item{[95A1]} M. Ambrosio et al., MACRO Coll. 
(Performance of the MACRO streamer tube
system in the search for magnetic monopoles) Astropart. Phys. 4 (1995) 33.
\vspace{-3.5 mm}
\item{[95A2]} M. Acciarri et al., L3 Coll. (Search for anomalous 
$Z \to \gamma \gamma \gamma $ events at LEP) Phys. Lett. B345 (1995) 609.
\vspace{-3.5 mm} 
\item{[95B1]} P. Bhattacharjee (Monopole annihilation and the highest energy 
cosmic rays) Phys. Rev. D51 (1995) 4079.
\vspace{-3.5 mm}
\item{[95C1]} C. Coriano and R. R. Parwani (The electric charge of a Dirac 
monopole at nonzero temperature) Phys. Lett. B363 (1995) 71.
\vspace{-3.5 mm}
\item{[95D1]} G. Dvali et al. (Symmetry nonrestoration at high temperature
and the monopole problem) Phys. Rev. Lett. 75 (1995) 4559.
\vspace{-3.5 mm}
\item{[95D2]} A. De Rujula (Effects of virtual monopoles) Nucl. Phys. B435 
(1995) 257.
\vspace{-3.5 mm}
\item{[95H1]} H. J. He and Z. Qiu (Inconsistency of QED in presence of Dirac 
monopoles) Z. Phys. C65 (1995) 175.
\vspace{-3.5 mm}
\item{[95J1]} H. Jeon and M. J. Longo (Search for magnetic monopoles trapped in
matter) Phys. Rev. Lett. 75 (1995) 1443.
\vspace{-3.5 mm}
\item{[95M1]} S. Mahan (Model for efficient monopole annihilation and 
baryogenesis) Mod. Phys. Lett. A10 (1995) 227.
\vspace{-3.5 mm}
\item{[95M2]} I. De Mitri et al. (Magnetic monopole trigger with streamer tubes
in the MACRO experiment at Gran Sasso) Nucl. Instrum. Meth. A360 (1995) 311.
\vspace{-3.5 mm}
\item{[95P1]} G. I. Poulis and P. J. Mulders (Is QED inconsistent in the 
presence of Dirac monopoles?) Z. Phys. C67 (1995) 181.
\vspace{-3.5 mm}
\item{[95V1]} V. M. Villalba (Exact solution of the Dirac equation for a 
Coulomb and a scalar potential in the presence of an Aharanov-Bohm and a 
magnetic monopole fields) J. Math. Phys. 36 (1995) 3332.

\item{[96B1]} G. Bimonte and G. Lozano (On symmetry nonrestoration at high 
temperature) Phys. Lett. B366 (1996) 248.
\vspace{-3.5 mm}
\item{[96B2]} G. Bimonte and G. Lozano (Can symmetry nonrestoration solve
the monopole problem?) Nucl. Phys. B460 (1996) 155.
\vspace{-3.5 mm}
\item{[96B3]} E. Bellotti, R. A. Carrigan, G. Giacomelli and N. Paver 
(Non-accelerator particle astrophysics) World Scientific (1996); 
G. Giacomelli (Magnetic monopole searches) Lectures given at Lake Louise 
Winter Institute on Particle Physics and Cosmology (1994).
\vspace{-3.5 mm}
\item{[96C1]} E. Comay (Charges, monopoles and duality relations)
Nuovo Cim. 110B (1996) 1347.
\vspace{-3.5 mm}
\item{[96C2]} S. Cecchini et al. (Calibration with relativistic and low 
velocity ions of a CR39 nuclear track detector) 
Nuovo Cim. 109A (1996) 1119.
\vspace{-3.5 mm}
\item{[96G1]} J. C. Le Guillou and F. A. Schaposnik  (On the electric charge
of monopoles at finite temperature) Phys. Lett. B383 (1996) 339.
\vspace{-3.5 mm}
\item{[96I1]} A. Y. Ignatev and G. C. Joschi  (Massive electrodynamics and the
magnetic monopoles) Phys. Rev. D53 (1996) 984.
\vspace{-3.5 mm}
\item{[96I2]} A. Y. Ignatev and G. C. Joschi  (Magnetic monopole and the finite 
photon mass: are they compatible?) Mod. Phys. Lett. A11 (1996) 2735.
\vspace{-3.5 mm}
\item{[96K1]} T. W. Kephart and T. J. Weiler  (Magnetic monopoles as the
highest energy cosmic ray primaries) Astropart. Phys. 4 (1996) 271.
\vspace{-3.5 mm}
\item{[96L1]} K. Lee, E. J. Weinberg and P. Yi  (Massive and massless monopoles 
with nonabelian Magnetic charges) Phys. Rev. D54 (1996) 6351.
\vspace{-3.5 mm}
\item{[96P1]} Particle Data Group (Review of particle physics)
Phys. Rev. D54 (1996) 1.
\vspace{-3.5 mm}
\item{[96S1]} N. Sakai  (Dynamics of gravitating magnetic
monopoles) Phys. Rev. D54 (1996) 1548.
\vspace{-3.5 mm}
\item{[96S2]} P. F. Spada  (Ricerca di monopoli magnetici col rivelatore MACRO
) Ph.D. Thesis, University of Bologna (1996).
\vspace{-3.5 mm}
\item{[96V1]} T. Vachaspati  (An attempt to construct the Standard Model with 
monopoles) Phys. Rev. Lett. 76 (1996) 188.
\vspace{-3.5 mm}
\item{[96W1]} T. J. Weiler and T. J. Kephart (Are we seeing magnetic
monopole cosmic rays at E$> 10^{20}$eV?) Nucl. Phys. Proc. Suppl. 51B (1996) 
218.

\item{[97A1]} S. P. Ahlen et al. (Energy loss of supermassive magnetic
monopoles and dyons in main sequence stars) Phys. Rev. D55 (1997) 6584.
\vspace{-3.5 mm}
\item{[97A2]} M.  Ambrosio et al., MACRO Coll. 
(The performance of MACRO liquid 
scintillator in the search for magnetic monopoles with $10^{-3} < \beta < 1$) 
Astropart. Phys. 6 (1997) 113.
\vspace{-3.5 mm}
\item{[97A3]} M.  Ambrosio et al., MACRO Coll. 
(Magnetic monopole search with the MACRO
detector at Gran Sasso) Phys. Lett. B406 (1997) 249.
\vspace{-3.5 mm}
\item{[97A4]} M.  Ambrosio et al., MACRO Coll.
(Magnetic monopole search with the MACRO detector at Gran Sasso) INFN/AE-97/19, 
Paper presented at the 25th ICRC, Durban, South Africa (1997). 
\vspace{-3.5 mm}
\item{[97A5]} M. Ambrosio et al., MACRO Coll.  
(Search for magnetic monopoles with the nuclear track 
detector of the MACRO experiment) Nucl. Tracks Radiat. Meas. 28 (1997) 297.  
\vspace{-3.5 mm}
\item{[97B1]} V. Berezinskii et al. (High-energy particles from monopoles 
connected by strings) Phys. Rev. D56 (1997) 2024.
\vspace{-3.5 mm}
\item{[97B2]} I. A. Belolaptikov et al. (The Baikal underwater neutrino 
telescope: design, performance and first results) Astropart. Phys. 7 (1997) 263.
\vspace{-3.5 mm}
\item{[97B3]} E. Bergshoeff et al. (Kaluza-Klein monopoles and gauged 
sigma models) Phys. Lett. B410 (1997) 131.
\vspace{-3.5 mm}
\item{[97B4]} D. Bak and H. Min (Radiation damping of a BPS monopole: an
implication to S duality) Phys. Rev. D56 (1997) 6665.
\vspace{-3.5 mm}
\item{[97B5]} P. Bhattacharjee, Q. Shafi, F. W. Stecker (TeV and superheavy
mass-scale particles from supersymmetric topological defects, the extragalactic
$\gamma-$ray background and the highest energy cosmic rays) hep-ph/9710533.
\vspace{-3.5 mm}  
\item{[97C1]} Y. M. Cho and D. Maison (Monopole configuration 
in the Weinberg-Salam model) 
Phys. Lett. B391 (1997) 360.
\vspace{-3.5 mm}
\item{[97C2]} I. Cho and A. Vilenkin (Space-time structure of an inflating 
global monopole)
Phys. Rev. D56 (1997) 7626.
\vspace{-3.5 mm}
\item{[97C3]} A. H. Chamseddine and M. S. Volkov (Non-abelian 
Bogomol'nyi-Prasad-Sommerfield monopoles in N = 4 gauged supergravity)
Phys. Rev. Lett. 79 (1997) 3343.
\vspace{-3.5 mm}
\item{[97C4]} Y. M. Cho and K. Kimm (Finite energy electroweak monopoles)
hep-th/9707038 (1997).
\vspace{-3.5 mm}
\item{[97D1]} G. Damm and W. Kerler (Monopoles and deconfinement transition 
in SU(2) lattice gauge theory)
Phys. Lett. B397 (1997) 216.
\vspace{-3.5 mm}
\item{[97F1]} R. Fantini (Ricerca di monopoli magnetici GUT con i tubi a 
streamer di MACRO)  Ph.D. Thesis, University of Bologna (1997).
\vspace{-3.5 mm}
\item{[97F2]} M. Faber et al. (Nambu-Jona-Lasinio version of magnetic monopole 
physics with dual Dirac strings) Z. Phys. C74 (1997) 721.
\vspace{-3.5 mm}
\item{[97G1]} A. Di Giacomo and M. Marthur (Magnetic monopoles, gauge  
invariant dynamical variables and Georgi Glashow model) 
Phys. Lett. B400 (1997) 129.
\vspace{-3.5 mm}
\item{[97G2]} G. Giacomelli and V. Popa (Search for magnetic monopoles and for 
nuclearites with the MACRO detector at Gran Sasso) DFUB 3/97, Invited paper at 
the 5th Topical Seminar, San Miniato, Italy, 21-25 April 1997.
\vspace{-3.5 mm}
\item{[97H1]} Y. D. He (Search for a Dirac magnetic monopole in high energy 
nucleus nucleus collisions) Phys. Rev. Lett. 79 (1997) 3134.
\vspace{-3.5 mm}
\item{[97I1]} P. Irwin (SU(3) monopoles and their fields)
Phys. Rev. D56 (1997) 5208.
\vspace{-3.5 mm}
\item{[97I2]} A. Y. Ignatev and G. C. Joschi (Dirac magnetic monopole and  
the discrete simmetries) hep-ph/9710553 (1997).
\vspace{-3.5 mm}
\item{[97K1]} S. V. Ketov (Solitons, monopoles and duality: from Sine-Gordon
to Seiberg-Witten) Fortsch. Phys. 45 (1997) 237.
\vspace{-3.5 mm}
\item{[97K2]} S. G. Kovalevich (The effective lagrangian of QED with a magnetic 
charge and dyon mass bounds) Phys. Rev. D55 (1997) 5807.
\vspace{-3.5 mm}
\item{[97L1]} H. Liu and T. Vachaspati (SU(5) monopoles and the dual standard
model) Phys. Rev. D56 (1997) 1300.
\vspace{-3.5 mm}
\item{[97L2]} O. Lebedev (A finite-size magnetic monopole in double-potential 
formalism) Mod. Phys. Lett. A12 (1997) 2203.
\vspace{-3.5 mm}
\item{[97R1]} H. Reinhardt (Resolution of Gauss' law in Yang-Mills theory by 
gauge invariant projection: topology and magnetic monopoles) 
Nucl. Phys. B503 (1997) 505.
\vspace{-3.5 mm}
\item{[97S1]} P. M. Sutcliffe (BPS monopoles)
Int. J. Mod. Phys. A12 (1997) 4663.
\vspace{-3.5 mm}
\item{[97S2]} C. Saclioglu and S. Nergiz (Seiberg-Witten monopole equations 
and Riemann surfaces) Nucl. Phys. B503 (1997) 675.
\vspace{-3.5 mm}
\item{[97T1]} G. F. Torres del Castillo and L. C. Cortes-Cuautli (Solution 
of the Dirac equation in the field of a magnetic monopole) 
J. Math. Phys. 38 (1997) 2996.
\vspace{-3.5 mm}
\item{[97T2]} E. Teo and C. Ting (Monopoles, vortices and kinks in the 
framework of noncommutative geometry) Phys. Rev. D56 (1997) 2291.
\vspace{-3.5 mm}
\item{[97V1]} A. S. Vshivtsev and K. G. Klimenko (Suppression of superheavy 
monopoles in Grand-Unification models) Phys. Atom. Nuclei 60 (1997) 1570.


\item{[98A1]} B. Abbott et al., D0 Coll. (A search for heavy pointlike 
Dirac monopoles) Phys. Rev. Lett. 81 (1998) 524, hep-ex/9803023.
\vspace{-3.5 mm}

\item{[98B1]}  B. Bajc, A. Riotto, G. Senjanovic (R-charge kills monopoles)
 Mod. Phys. Lett. A13 (1998) 2955, hep-ph/9803438.
\vspace{-3.5 mm}

\item{[98B2]} O. Bergman, B. Kol (String webs and 1/4 BPS monopoles)
  Nucl. Phys. B536 (1998) 149, hep-th/9804160.
\vspace{-3.5 mm}

\item{[98B3]} E. A. Bergshoeff (Kaluza-Klein monopoles and gauged sigma models)
Nucl. Phys. Proc. Suppl. 68 (1998) 355.
\vspace{-3.5 mm}

\item{[98B4]} R. Bielawski (Asymptotic metrics for SU(N) monopoles 
with maximal symmetry breaking) Commun. Math. Phys. 199 
(1998) 297, hep-th/9801092.
\vspace{-3.5 mm}

\item{[98D1]} U. H. Danielsson, A. P. Polychronakos (Quarks, monopoles 
and dyons at large N) Phys. Lett. B434 (1998) 294, hep-th/9804141.
\vspace{-3.5 mm}

\item{[98D2]} T. Dereli, M. Tekmen (Wu-Yang monopoles and nonabelian 
Seiberg-Witten equations) Mod. Phys. Lett. A13 (1998) 1803, hep-th/9806031.
\vspace{-3.5 mm}

\item{[98D3]} J. Derkaoui, G. Giacomelli, T. Lari, A. Margiotta, M. Ouchrif,
L. Patrizii, V. Popa and V. Togo (Energy losses of magnetic monopoles and 
of dyons in the earth) Astropart. Phys. 9 (1998) 173.
\vspace{-3.5 mm}

\item{[98D4]} A. Di Giacomo, M. Mathur (Abelianization of SU(N) gauge 
theory with gauge invariant dynamical variables and magnetic monopoles)
 Nucl. Phys. B531 (1998) 302, hep-th/9802050. \par
\vspace{-3.5 mm}

\item{[98E1]} E. Eyras, B. Janssen, Y. Lozano (Five-branes, K K monopoles 
and T duality) Nucl. Phys. B531 (1998) 275, hep-th/9806169.
\vspace{-3.5 mm}

\item{[98F1]} C. Ford, U. G. Mitreuter, T. Tok, A. Wipf and J. M. Pawlowski 
(Monopoles, Polyakov loops and gauge fixing on the torus) Annals Phys. 
269 (1998) 26, hep-th/9802191.
\vspace{-3.5 mm}

\item{[98G1]} L. Gamberg, G. R. Kalbfleisch, K. A. Milton
(Difficulties with photonic searches for magnetic monopoles) hep-ph/9805365.
\vspace{-3.5 mm}

\item{[98G2]} G. Giacomelli, L. Patrizii (Magnetic monopoles) Proceedings of
the 5$^{th}$ School on Non-accelerator particle astrophysics 
(Trieste, Italy, 1998) 285-297, hep-ex/0002032.
\vspace{-3.5 mm}

\item{[98G3]}  Yu. P. Goncharov (U(N) monopoles on kerr black hole and 
its entropy) Mod. Phys. Lett. A13 (1998) 1495, gr-qc/9806020.
\vspace{-3.5 mm}

\item{[98G4]} M. Grady (Gauge invariant SO(3) - Z2 monopoles as possible 
source of confinement in SU(2) lattice gauge theory) hep-lat/9806024.
\vspace{-3.5 mm}

\item{[98G5]} I. F. Ginzburg, A. Schiller (Search for a heavy magnetic monopole
at the Fermilab Tevatron and CERN LHC) hep-ph/9802310.
\vspace{-3.5 mm}

\item{[98G6]} G. Giacomelli et al., Magnetic monopole bibliography, DFUB 98-9.
\vspace{-3.5 mm}
 
\item{[98H1]} Y. D. He (Can track etch detector CR39 record low velocity GUT
 magnetic monopoles?) Phys. Rev. D57 (1998) 3182.
\vspace{-3.5 mm}

\item{[98I1]} E. M. Ilgenfritz, H. Markum, M. Mueller-Preussker, S. Thurner 
(Action and topological density carried by abelian monopoles in finite
temperature pure SU(2) gauge theory: an analysis using RG smoothing) Phys. 
 Rev. D58:094502 (1998), hep-lat/9801040.
\vspace{-3.5 mm}

\item{[98J1]} O. Jahn, F. Lenz (Structure and dynamics of monopoles 
in axial gauge QCD) Phys. Rev. D58:085006 (1998), hep-th/9803177.
\vspace{-3.5 mm}

\item{[98K1]} B. Kleihaus, J. Kunz, D. H. Tchrakian (Interaction energy 
of `t Hooft-Polyakov monopoles) Mod. Phys. Lett. A13 (1998) 2523,
 hep-th/9804192
\vspace{-3.5 mm}

\item{[98K2]} T. C. Kraan, P. van Baal (Monopole constituents inside
SU(n) calorons) hep-th/9806034. 
\vspace{-3.5 mm}

\item{[98L1]} K. Y. Lee, C. Lu (SU(2) calorons and magnetic monopoles)
 Phys. Rev. D58:025011 (1998), hep-th/9802108.
\vspace{-3.5 mm}

\item{[98L2]} K. Lee (Instantons and magnetic monopoles on $R^3 \times 
S^1$ with 
arbitrary simple gauge groups) Phys. Lett. B426 (1998) 323, hep-th/9802012.
\vspace{-3.5 mm}

\item{[98S1]} S. Sasaki, O. Miyamura (Lattice study of U(A)(1) 
anomaly: the role of QCD monopoles) Phys. Lett. B443 
(1998) 331, hep-lat/9810039.
\vspace{-3.5 mm}

\item{[98T1]} B. Tekin, K. Saririan, Y. Hosotani
(Complex monopoles in YM + Chern-Simons theory in 3 dimensions)
Talk given at the 10th International Seminar on High-Energy Physics (Quarks
98), Suzdal, Russia (1998), hep-th/9808057.


\item{[99A1]} M. Ambrosio et al., MACRO Coll. (Search for supermassive 
magnetic monopoles with the MACRO detector at
the Gran Sasso laboratory) Nucl. Phys. Proc. Suppl. 70 (1999) 466.
\vspace{-3.5 mm}

\item{[99A2]} M. Ambrosio et al., MACRO Coll. (Search for magnetic 
monopoles with MACRO) Salt Lake City 1999, Int. Cosmic Ray 
Conf., vol. 2 332-335, hep-ex/9905023.
\vspace{-3.5 mm}

\item{[99A3]} M. Ambrosio et al., MACRO Coll. (Search for magnetic monopoles 
and nuclearites with the MACRO nuclear track detector) 
Radiat. Meas. 31 (1999) 605.
\vspace{-3.5 mm}

\item{[99A4]} M. Ambrosio et al., MACRO Coll. (Search for magnetic monopoles 
with nuclear track detectors) Talk given at 6th Topical Seminar on Neutrino 
and AstroParticle Physics, San Miniato, Italy (1999), hep-ex/9909012.
\vspace{-3.5 mm}

\item{[99B1]} P. van Baal (Instantons versus monopoles) hep-th/9912035.
\vspace{-3.5 mm}

\item{[99B2]} V. A. Balkanov (Search for relativistic monopoles with 
the Baikal neutrino telescope) Salt Lake City 1999, Int. Cosmic 
Ray Conf., vol. 2 340-343.
\vspace{-3.5 mm}

\item{[99C1]} B. Chen, H. Itoyama, H. Kihara (Nonabelian monopoles from 
matrices: seeds of the space-time structure) hep-th/9909075.
\vspace{-3.5 mm}

\item{[99C2]} M. N. Chernodub, M. I. Polikarpov, A. I. Veselov, M. A. Zubkov 
(Aharonov-Bohm effect, center monopoles and center vortices in SU(2)
lattice gluodynamics) Nucl. Phys. Proc. Suppl. 73 (1999) 575, hep-lat/9809158.
\vspace{-3.5 mm}

\item{[99D1]} N. M. Davies, T. J. Hollowood, V. V. Khoze, M. P. Mattis 
(Gluino condensate and magnetic monopoles in supersymmetric gluodynamics)
 Nucl. Phys. B559 (1999) 123,  hep-th/9905015.
\vspace{-3.5 mm}

\item{[99D2]} J. Derkaoui , G. Giacomelli, T. Lari, G. Mandrioli, M. 
Ouchrif, L. Patrizii, V. Popa (Energy losses of magnetic monopoles 
and dyons in scintillators, streamer tubes and nuclear track detectors)
 Astropart. Phys. 10 (1999) 339.
\vspace{-3.5 mm}

\item{[99F1]} P. M. N. Feehan (Generic metrics, irreducible rank one 
PU(2) monopoles, and transversality) submitted to Commun. Anal. Geom.,
 math.dg/9809001.
\vspace{-3.5 mm}

\item{[99F2]} K. Freese, E. Krasteva (A bound on the flux of magnetic 
monopoles from catalysis of nucleon decay in white 
dwarfs) Phys. Rev. D59:063007 (1999), astro-ph/9804148.
\vspace{-3.5 mm}

\item{[99G1]} G. Gabadadze, Z. Kakushadze (A remark on Witten effect 
for QCD monopoles in matrix quantum mechanics) Mod. Phys. Lett. A14 (1999)
2151, hep-th/9908039.
\vspace{-3.5 mm}

\item{[99G2]} L. Gamberg, G. R. Kalbfleisch, K. A. Milton (Direct and 
indirect searches for low mass magnetic monopoles) hep-ph/9906526.
\vspace{-3.5 mm}

\item{[99G3]} J. P. Gauntlett, C. Koehl, D. Mateos, P. K. Townsend, M. 
Zamaklar (Finite energy Dirac-Born-Infeld monopoles and string junctions)
Phys. Rev. D60:045004 (1999), hep-th/9903156. \par
\vspace{-3.5 mm}

\item{[99G4]} A. S. Goldhaber (Dual confinement of grand unified monopoles?)
 Phys. Rep. 315 (1999) 83, hep-th/9905208.
\vspace{-3.5 mm}

\item{[99G5]} N. Grandi, E. F. Moreno, F. A. Schaposnik (Monopoles in 
nonabelian Dirac-Born-Infeld theory) Phys. Rev. D59:125014 
(1999), hep-th/9901073
\vspace{-3.5 mm}

\item{[99H1]} A. Hanany (Monopoles in string theory)
JHEP 9912:014 (1999), hep-th/9911113.
\vspace{-3.5 mm}

\item{[99H2]} Y. Hosotani, K. Saririan, B. Tekin (Complex monopoles and 
gribov copies) Talk given at 3rd Workshop on Continuous Advances in QCD 
(QCD 98), Minneapolis, MN (1998),  hep-th/9808105.
\vspace{-3.5 mm}

\item{[99I1]} H. Ichie, H. Suganuma (Monopoles and gluon fields in QCD 
in the maximally abelian gauge), hep-lat/9808054.
\vspace{-3.5 mm}

\item{[99I2]} T. Ioannidou, P. M. Sutcliffe (Nonbogomolny SU(N) BPS monopoles)
 Phys. Rev. D60:105009 (1999), hep-th/9905169.
\vspace{-3.5 mm}

\item{[99K1]} T. C. Kraan, P. van Baal (Constituent monopoles without 
gauge fixing) Nucl. Phys. Proc. Suppl. 73 (1999) 554, hep-lat/9808015.
\vspace{-3.5 mm}

\item{[99L1]} K. Lee  (Sheets of BPS monopoles and instantons with arbitrary 
simple gauge group) Phys. Lett. B445 (1999) 387, hep-th/9810110.
\vspace{-3.5 mm}

\item{[99L2]} K. Lee (Massless monopoles and multipronged strings)
Phys. Lett. B458 (1999) 53, hep-th/9903095.
\vspace{-3.5 mm}

\item{[99L3]} A. Lue, E. J. Weinberg (Magnetic monopoles near the black 
hole threshold) Phys. Rev. D60:084025 (1999), hep-th/9905223.
\vspace{-3.5 mm}

\item{[99L4]} M. J. Lewis, K. Freese, G. Tarl\`e (Protogalactic 
extension of the Parker bound) astro-ph/9911095.   
\vspace{-3.5 mm}

\item{[99N1]} P. Niessen et al., AMANDA Coll. (Search for relativistic 
monopoles with the AMANDA detector) Salt Lake City 1999, Int. Cosmic 
Ray Conf., vol. 2 344-347.
\vspace{-3.5 mm}

\item{[99P1]} L. Pogosian, T. Vachaspati (Interaction of magnetic 
monopoles and domain walls)  hep-ph/9909543.
\vspace{-3.5 mm}

\item{[99Q1]} M. Quandt, H. Reinhardt, A. Schafke (Magnetic monopoles and 
topology of Yang-Mills theory in Polyakov gauge) Phys. Lett. B446 (1999) 290,
 hep-th/9810088.
\vspace{-3.5 mm}

\item{[99R1]} H. Reinhardt, M. Engelhardt, K. Langfeld, M. Quandt, A. Schafke
(Magnetic monopoles, center vortices, confinement and topology of gauge
fields) hep-th/9911145.
\vspace{-3.5 mm}

\item{[99S1]} D. Sassi Thober (The monopoles in the structure of the electron)
 hep-ph/9906377.
\vspace{-3.5 mm}

\item{[99T1]} B. Tekin, K. Saririan, Y. Hosotani
(Complex monopoles in the Georgi-Glashow-Chern-Simons model)
Nucl. Phys. B539 (1999) 720, hep-th/9808045.
\vspace{-3.5 mm}

\item{[99T2]}  A. Teleman (Moduli spaces of PU(2) monopoles)
 submitted to Asian J.Math,  math.dg/9906163.
\vspace{-3.5 mm}

\item{[99T3]} P. K. Tripathy (Gravitating monopoles and black holes in 
Einstein-Born-Infeld-Higgs model) Phys. Lett. B458 (1999) 252, hep-th/9904186.
\vspace{-3.5 mm}

\item{[99W1]} E. J. Weinberg (Massive and massless monopoles and duality)
 hep-th/9908095.


\item{[00B1]} D. Bakari et al., (Magnetic monopoles, nuclearites, Q-balls: a 
qualitative picture) hep-ex/0004019.
\vspace{-3.5 mm}

\item{[00B2]} D. Bakari et al., SLIM Coll. (Search for ``light" magnetic 
monopoles) hep-ex/0003028.
\vspace{-3.5 mm}

\item{[00B3]}  V. A. Balkanov et al., (Search for fast magnetic
    monopoles in a deep-sea experiment at Lake Baikal)
    Bull. Russ. Acad. Sci. Phys. 63 (1999) 485, Izv. Ross.
    Akad. Nauk, Fiz. 63 (1999) 598.
\vspace{-3.5 mm}

\item{[00D1]} N. M. Davies, V. V. Khoze (On Affleck-Dine-Seiberg superpotential
 and magnetic monopoles in supersymmetric QCD) 
JHEP 0001:015 (2000), hep-th/9911112.
\vspace{-3.5 mm}

\item{[00D2]} Zh. Dzhilkibaev
 (Search for fast monopoles in the Baikal experiment)
Zeuthen 1998, Simulation and analysis methods for large neutrino 
telescopes (1998) 364.
\vspace{-3.5 mm}

\item{[00G1]} L. Gamberg, K. A. Milton (Eikonal scattering
    of monopoles and dyons in dual QED) hep-ph/0005016.
\vspace{-3.5 mm}

\item{[00H1]}  Y. Hosotani, J. Bjoraker (Monopoles and
    dyons in the pure Einstein-Yang-Mills theory) gr-qc/0001105.
\vspace{-3.5 mm}

\item{[00J1]} O. Jahn (Instantons and monopoles in general abelian gauges)
J. Phys. A33 (2000) 2997, hep-th/9909004.
\vspace{-3.5 mm}

\item{[00J2]}  R. Jeannerot, S. Khalil, G. Lazarides, Q. Shafi (Inflation 
and monopoles in supersymmetric SU(4)C X SU(2)(L) X SU(2)(R)) hep-ph/0002151.
\vspace{-3.5 mm}

\item{[00K1]} G. R. Kalbfleisch et al., (Improved experimental limits on
    the production of magnetic monopoles) hep-ex/0005005.
\vspace{-3.5 mm}

\item{[00K2]} B. Kol, M. Kroyter, (On the spatial structure of
    monopoles) hep-th/0002118.
\vspace{-3.5 mm}

\item{[00L1]} N. F. Lepora, (On the spectrum and representation
    theory of nonabelian monopoles)  hep-th/0002163.
\vspace{-3.5 mm}

\item{[00L2]} S. L. Liebling (Static gravitational global monopoles)
Phys. Rev. D61:024030 (2000), gr-qc/9906014.
\vspace{-3.5 mm}

\item{[00M1]} D. Maison (Gravitational global monopoles with horizons) 
 gr-qc/9912100.
\vspace{-3.5 mm}

\item{[00W1]} S. D. Wick, T. W. Kephart, T. J. Weiler, P. L. Biermann  
 (Signatures for a cosmic flux of magnetic monopoles) submitted to
Astropart. Phys., astro-ph/0001233.
\vspace{-3.5 mm}

\vspace{2cm}

Other MM bibliographies can be found 
in: [73S1], [77C1], [80R1], [82C4], [84G1], [94G1], [98G6].

\newpage

\begin{figure}
\begin{center}
        \mbox{ \epsfysize=7.6cm
            \epsffile{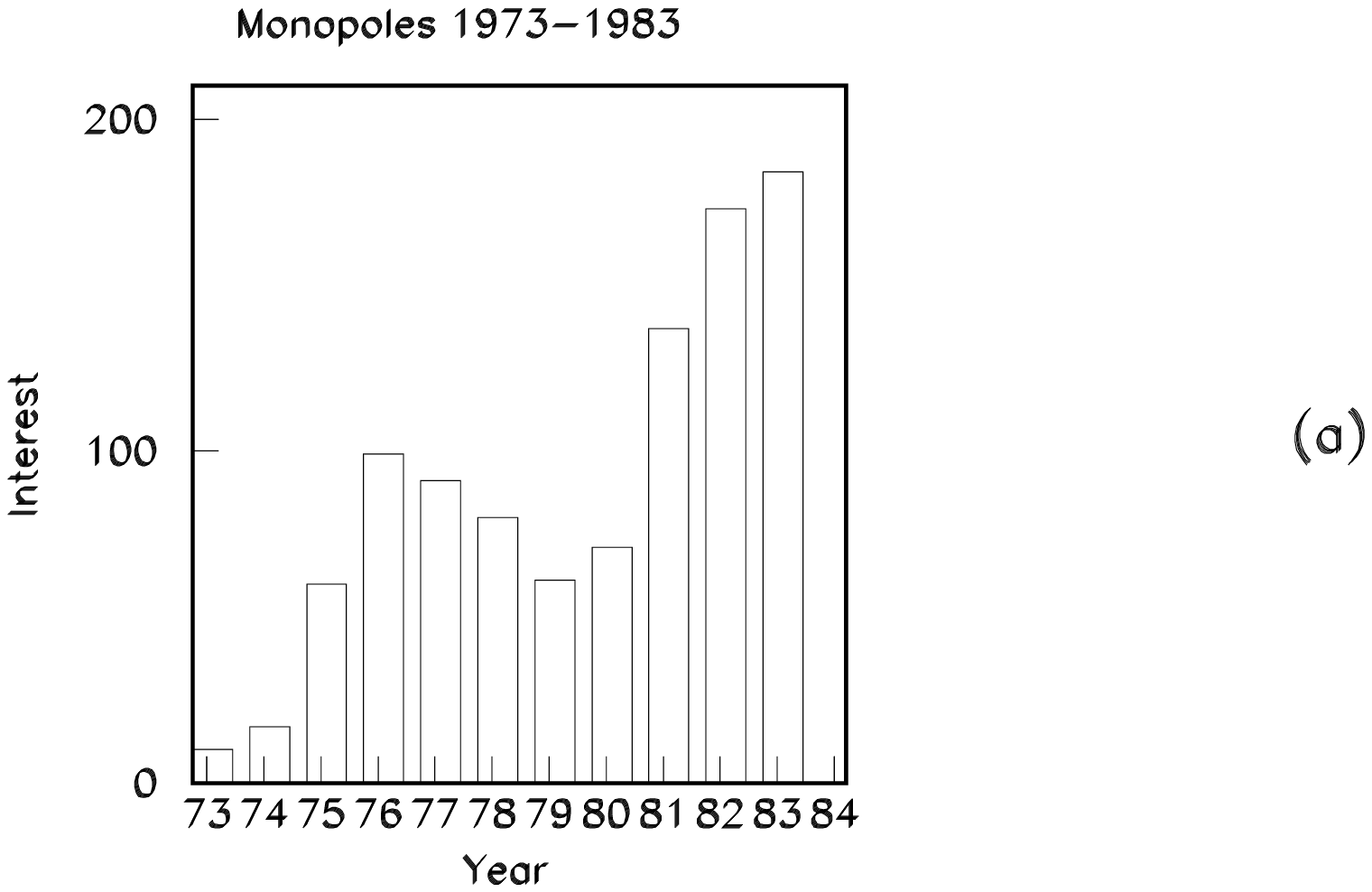}}
        \mbox{ \epsfysize=12.5cm
            \epsffile{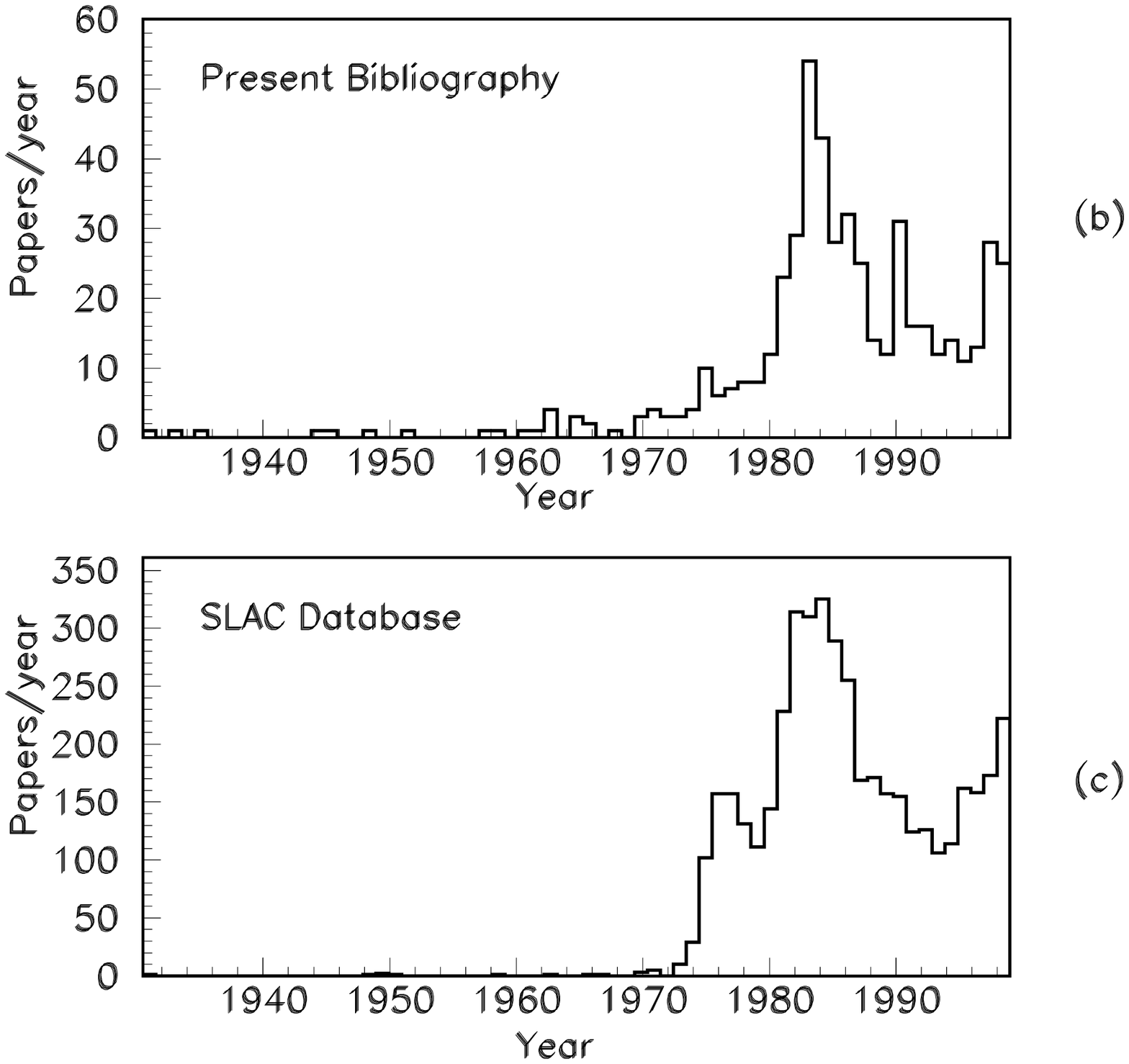}}
\caption {\small (a) The upper hystogram shows the yearly 
number of papers on MMs published from 
1973 till 1983 [83P1].
 (b) The middle hystogram shows the number of papers of the present 
bibliography as a function of the year of publication. (c) The lower 
hystogram shows the yearly number of papers found in the SLAC database 
with the search command {\em find title monopole\# or title dyon\# or 
keyword magnetic monopole} as a function of the year 
of entry in the database.}  
\end{center}
\end{figure}

\newpage

\begin{figure}
\begin{center}
        \mbox{ \epsfysize=12cm
            \epsffile{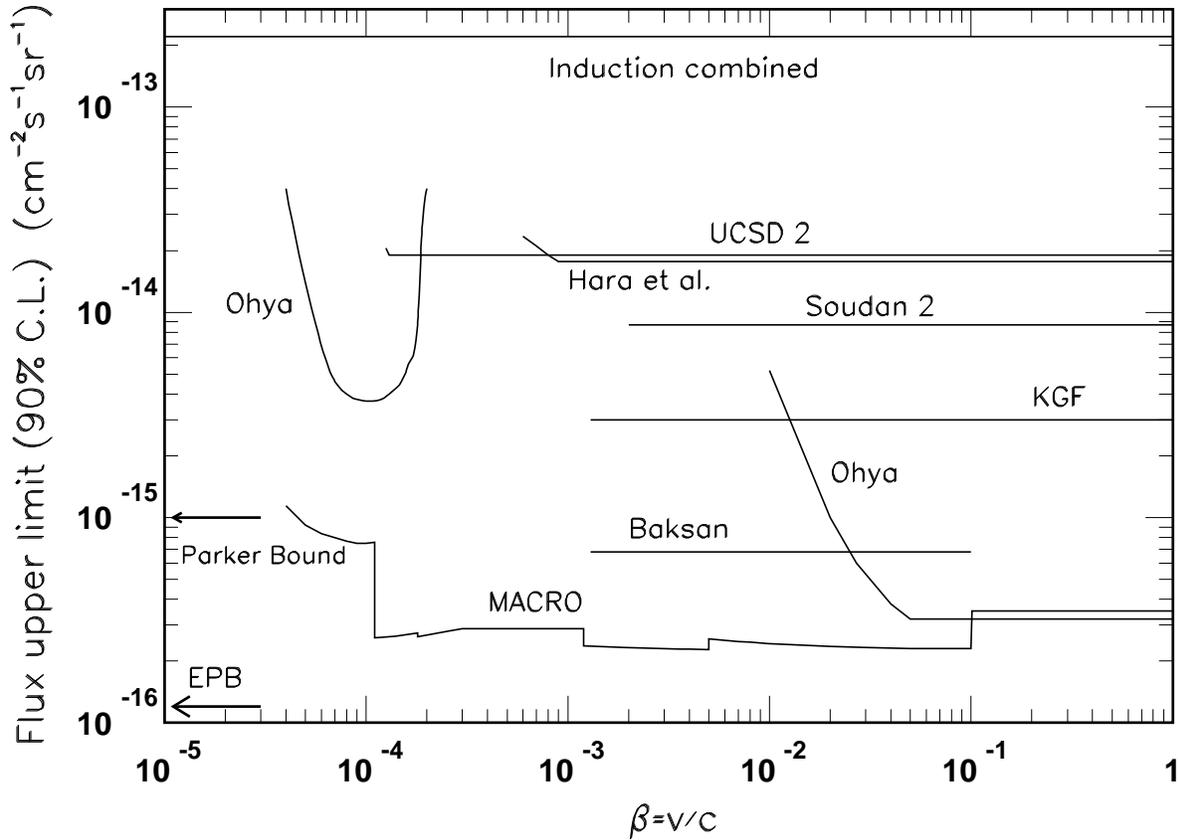}}
\caption{\small  90\% C.L. upper limits on an isotropic 
flux of $g=g_{D}$, massive magnetic 
monopoles in the cosmic radiation, assuming a catalysis cross section smaller 
than a few mb [97A4], [97G2]. The limits are now regularly updated.}
\end{center}
\end{figure}

\end{description}

\end{document}